\documentclass[10pt, journal, final]{IEEEtran}

\ifCLASSINFOpdf
\else
\fi

\usepackage{graphicx}
\usepackage[cmex10]{amsmath}
\usepackage{hyperref}
\usepackage{amsfonts}
\usepackage{dblfloatfix}
\usepackage{cite}
\usepackage{dsfont}
\usepackage{bm}
\usepackage{amssymb}
\usepackage{array}
\usepackage[ruled,vlined]{algorithm2e}
\usepackage{mathrsfs}

\usepackage{subfigure}
\usepackage{graphics}
\usepackage{epsfig}
\usepackage{epstopdf}
\usepackage{float}
\usepackage{extarrows}
\usepackage{booktabs}
\usepackage{xcolor}

\newtheorem{remark}{Remark}

\begin{document}

\title{Cooperative Multi-Cell Massive Access with Temporally Correlated Activity}
%

\author{
Weifeng~Zhu,~\IEEEmembership{Graduate Student Member,~IEEE}, Meixia~Tao,~\IEEEmembership{Fellow,~IEEE}, \\ Xiaojun~Yuan,~\IEEEmembership{Senior Member,~IEEE}, Fan Xu, and Yunfeng~Guan

\thanks{W. Zhu, M. Tao, and Y. Guan are with the Department of Electronic Engineering and the Cooperative Medianet Innovation Center (CMIC), Shanghai Jiao Tong University, Shanghai 200240, China (e-mail: \{wf.zhu, mxtao, yfguan69\}@sjtu.edu.cn).}
\thanks{X. Yuan is with the National Key Laboratory on Wireless Communications, University of Electronic Science and Technology of China, Chengdu 610000, China (e-mail: xjyuan@uestc.edu.cn).}
\thanks{F. Xu is with Peng Cheng Laboratory, Shenzhen, Guangdong, 518000, China (e-mail: xuf02@pcl.ac.cn)}

}
\maketitle

\vspace{-1.9cm}

\begin{abstract}

This paper investigates the problem of activity detection and channel estimation in cooperative multi-cell massive access systems with temporally correlated activity, where all access points (APs) are connected to a central unit via fronthaul links.
We propose to perform user-centric AP cooperation for computation burden alleviation and introduce a generalized sliding-window detection strategy for fully exploiting the temporal correlation in activity.
By establishing the probabilistic model associated with the factor graph representation, we propose a scalable Dynamic Compressed Sensing-based Multiple Measurement Vector Generalized Approximate Message Passing (DCS-MMV-GAMP) algorithm from the perspective of Bayesian inference. Therein, the activity likelihood is refined by performing standard message passing among the activities in the spatial-temporal domain and GAMP is employed for efficient channel estimation.
Furthermore, we develop two schemes of quantize-and-forward (QF) and detect-and-forward (DF) based on DCS-MMV-GAMP for the finite-fronthaul-capacity scenario, which are extensively evaluated under various system limits.
Numerical results verify the significant superiority of the proposed approach over the benchmarks. Moreover, it is revealed that QF can usually realize superior performance when the antenna number is small, whereas DF shifts to be preferable with limited fronthaul capacity if the large-scale antenna arrays are equipped.

\end{abstract}
\begin{IEEEkeywords}
Multi-cell massive access, temporally correlated activity, cooperative activity detection, channel estimation, generalized approximate message passing (GAMP).
\end{IEEEkeywords}

\section{Introduction}\label{sec:introduction}

Massive machine-type communication (mMTC), as one of the vital services in the sixth generation (6G) cellular networks, can provide seamless massive connectivity for ubiquitous Internet of Things (IoT) applications \cite{Dawy_2017_WCM, Chen_2021_JSAC}. Due to the extremely large quantity of IoT devices, the conventional grant-based random access protocols can result in severe collision and very large signaling overhead \cite{Liu_2018_SPM}. By contrast, the recently emerged grant-free (GF) random access mechanism enables each active device to transmit pilot and data information to its access point (AP) directly without experiencing the elaborated handshaking procedure. To this end, the GF random access scheme becomes a promising candidate that can not only save the coordination overhead but also realize highly reliable connections \cite{Chen_2021_JSAC, Liu_2018_SPM, Shahab_2020_CST}. In GF schemes, each user is typically pre-assigned a unique pilot sequence for identification and channel estimation. However, due to the massive number of devices and yet limited time-frequency resources, pilot sequences cannot be mutually orthogonal. As such, a major challenge in massive access systems is the activity detection (AD) and channel estimation (CE) with non-orthogonal pilots. While many existing algorithms have been proposed in the literature for single-cell massive-access systems, this work aims to address the AD and CE problem in multi-cell massive access systems.

This paper adopts the cloud radio access network (C-RAN) architecture \cite{Checko_2015_CST} in multi-cell massive access systems for cooperative signal processing. 
In C-RAN, the baseband units from multiple APs are pooled into a central unit (CU) for improving the network capacity and energy efficiency \cite{Checko_2015_CST}.
In addition to AP cooperation, we also exploit the temporally correlated device activity to provide high-accuracy AD and CE.
The temporal correlation arises from the fact that the activated users often have a large probability to keep transmitting the information to the AP over multiple consecutive frames.
The design of the joint AD and CE approach that leverages both the AP cooperation and temporally correlated activity for multi-cell networks is very challenging.
Furthermore, the fronthaul links are often capacity-limited in practice, due to the imperfect hardware equipment. The exchanged contents between the APs and the CU for cooperation should be carefully compressed, which further increases the difficulty of cooperative AD and CE scheme design in multi-cell networks.

\subsection{Related Works}

The most prevalent approach for AD and CE in traditional single-cell massive-access systems is the compressed sensing (CS)-based methods which can well exploit the sparse nature of the activity of IoT devices \cite{Chen_2018_TSP, Liu_2018_TSP, Senel_2018_TCOM, Ke_2020_TSP, Zou_2020_SPL}. In particular, the Bayesian methods based on approximate message passing (AMP) have demonstrated attractive performance in \cite{Chen_2018_TSP, Liu_2018_TSP, Senel_2018_TCOM}.
Another promising approach is the covariance-based scheme which is suitable for massive MIMO systems where each AP is equipped with a huge number of antennas. The covariance-based scheme utilize the sample covariance matrix of received pilot signals to perform AD only without estimating the channels. Compared with the CS-based methods, covariance-based schemes can realize much better AD performance under fixed pilot length \cite{Fengler_2021_TIT, Chen_2021_TIT}.

Recently, the AD problem in cooperative multi-cell networks has attracted many attentions in \cite{Xu_2015_ICC, Utkovski_2017_SPL, Chen_2019_TWC, Ke_2021_JSAC, Shao_2020_TSP, JDD_2021_TWC, Chen_2021_TSP}. In specific, the work \cite{Xu_2015_ICC} proposes a modified Bayesian CS algorithm to detect the activities of all users in the C-RAN by processing the signals received at each AP jointly, where the fronthaul link has unlimited capacity. By considering limited fronthaul capacity, the work \cite{Utkovski_2017_SPL} compares two fronthauling schemes, namely, quantize-and-forward (QF) and detect-and-forward (DF). In QF, each AP first quantizes its received signals and then sends them to the CU for centralized activity detection. In DF, each AP performs distributed activity detection and then forwards the quantized log-likelihood ratios (LLRs) of the detection results to the CU for the final decision.
Afterward, assuming the total number of antennas in the system is fixed, the work \cite{Chen_2019_TWC} validates that the cooperative MIMO can improve the detection performance of cell-edge users compared with the non-cooperative massive MIMO that treats inter-cell inference as noise. Meanwhile, the work \cite{Ke_2021_JSAC} develops two processing paradigms of cloud computing and edge computing under QF for cell-free massive MIMO systems.
Note that the cooperative AD schemes with both QF and DF fronthauling in \cite{Utkovski_2017_SPL} are limited to the single-antenna scenario, while the multi-antenna scenario as considered in \cite{Chen_2019_TWC, Ke_2021_JSAC} only concentrates on either QF or DF, but not both. The direct performance comparison between QF and DF with multi-antenna AP is still missing in the literature.
To improve the performance of the CS-based methods in \cite{Xu_2015_ICC, Chen_2019_TWC, Ke_2021_JSAC}, the works \cite{Shao_2020_TSP, JDD_2021_TWC} propose the multi-cell covariance-based algorithms for cooperative AD with unlimited fronthaul capacity. Then the quantization schemes for the multi-cell covariance methods in the scenario with limited fronthaul capacity are also studied in \cite{Chen_2021_TSP}. However, these covariance methods are only applicable to the system with massive MIMO-equipped AP.


The aforementioned works \cite{Chen_2018_TSP, Liu_2018_TSP, Senel_2018_TCOM, Ke_2020_TSP, Zou_2020_SPL, Fengler_2021_TIT, Chen_2021_TIT, Xu_2015_ICC, Utkovski_2017_SPL, Chen_2019_TWC, Ke_2021_JSAC, Shao_2020_TSP, JDD_2021_TWC, Chen_2021_TSP} all perform AD independently in each frame.
To take advantage of the temporal correlation in device activities, the works \cite{Jiang_2021_TWC, Wang_2021_ISIT,Zhu_2022_WCL,Zhu_2023_TCOM} propose to solve the joint AD and CE problem from the dynamic compressed sensing (DCS) perspective \cite{Ziniel_2013_TSP}. In the works \cite{Jiang_2021_TWC} and \cite{Wang_2021_ISIT}, the historical estimation information is utilized to improve the AD and CE performance in the current frame. In particular, a sequential AMP (S-AMP) algorithm is proposed in \cite{Jiang_2021_TWC} based on the factor graph, and a side information-aided multiple measurement vector AMP (SI-aided MMV-AMP) algorithm is proposed in \cite{Wang_2021_ISIT} by identifying the SI from the estimation results in the last frame. Considering that the activity in the current frame is correlated with those in both the previous and following frames, the work \cite{Zhu_2022_WCL} makes use of the double-sided SI from both the previous frame and the next frame and proposes an AMP-SI algorithm with vector shrinkage function.
In contrast to the frame-by-frame detection \cite{Jiang_2021_TWC, Wang_2021_ISIT, Zhu_2022_WCL}, the work \cite{Zhu_2023_TCOM} proposes to detect the active users and estimate their channels in several consecutive frames simultaneously in a block-by-block manner. This strategy improves the performance with neglectable complexity increase.
By assuming that users have dynamic activities at each instant of the data phase, the work \cite{Yuan_2020_TCOM} adopts the spread data sequence and proposes to perform data detection along with activity tracking and CE based on an expectation maximization and hybrid message passing algorithm.
Note that these works \cite{Jiang_2021_TWC, Wang_2021_ISIT, Zhu_2023_TCOM, Yuan_2020_TCOM, Zhu_2022_WCL} only concentrate on the single-cell system without AP cooperation. Moreover, these AMP-based algorithms \cite{Jiang_2021_TWC, Wang_2021_ISIT, Zhu_2022_WCL} cannot well accommodate the signal quantization in the case with limited fronthaul capacity.
To the best of our knowledge, the AD and CE problem by jointly considering the temporally correlated activity and AP cooperation in multi-cell networks has not been studied to date.


\subsection{Main Contributions}

This paper considers the joint AD and CE problem for cooperative multi-cell temporally correlated massive access, where the C-RAN architecture with both infinite and finite fronthaul link capacities is adopted. First, we introduce a generalized sliding-window strategy for detection. Then, we establish the probabilistic model with the graphical representation to describe the statistical dependencies of the received signals. By formulating the joint AD and CE problem as a Bayesian inference problem, a computationally efficient dynamic compressed sensing-based multiple measurement vector generalized AMP (DCS-MMV-GAMP) algorithm is subsequently proposed, which can make full use of the spatial-temporal correlation in user activities.
Furthermore, two cooperative AD schemes based on DCS-MMV-GAMP are developed for the finite-fronthaul-capacity scenario.
The main contributions and results of this work are summarized as follows:
\begin{enumerate}
  \item

  We propose to formulate the AD and CE problem in multi-cell massive access with temporally correlated activity from the DCS perspective.
  The user-centric AP cooperation is applied for efficient joint signal processing in large-scale multi-cell networks. Compared with \cite{Chen_2019_TWC, Ke_2021_JSAC}, the detection AP set for each user is decided by the statistical channel information, which can achieve a better tradeoff between the fronthaul transmission cost and the detection performance.
  To balance the performance and the detection latency in exploiting the temporally correlated activity, we propose a generalized sliding-window detection strategy that can involve the existing sliding-window detection strategy \cite{Jiang_2021_TWC,Wang_2021_ISIT,Zhu_2023_TCOM,Zhu_2022_WCL} as its special cases.
  \item

  Based on the probabilistic model, we build the associated factor graph and propose a DCS-MMV-GAMP algorithm under the hybrid GAMP (HyGAMP) framework, which can approximately achieve Bayes-optimal performance with computational simplicity. In particular, the user-centric AP cooperation strategy can simplify the factor graph, so that the computational complexity of DCS-MMV-GAMP can be greatly reduced in large-scale systems. In contrast to the heuristic activity refinement design in \cite{Chen_2019_TWC,Ke_2021_JSAC,Jiang_2021_TWC,Wang_2021_ISIT,Zhu_2022_WCL}, we propose to update the activity likelihood based on the standard MP rule \cite{Ksch_2001_TIT} with performance enhancement. Compared with \cite{Chen_2019_TWC,Ke_2021_JSAC,Jiang_2021_TWC,Wang_2021_ISIT,Zhu_2023_TCOM,Zhu_2022_WCL}, the activity likelihood is updated by combining the messages from both the associate APs and the adjacent frames under the generalized sliding-window strategy, which further improves the performance.
  \item
  To deal with the fronthaul capacity limit in practical systems, we further develop QF and DF based on DCS-MMV-GAMP for cooperative AD, depending on the function split of the C-RAN. 
  Particularly, DCS-MMV-GAMP is modified to take the signal quantization into consideration for QF, which is not accounted in the conventional AMP-based algorithms\cite{Chen_2018_TSP, Wang_2021_ISIT, Ziniel_2013_TSP}.
  Extensive numerical results are also provided to compare the performance of these two schemes.
  It is worth noting that even limited fronthaul capacity can support QF to have superior performance to DF when there are only a few antennas at each AP. Whereas, DF can support significantly higher-quality AD than QF under the limited capacity scenario if the large-scale antenna array is equipped.
\end{enumerate}

\subsection{Organizations and Notations}

The remaining part of this paper is organized as follows.
Section \ref{sec:system_model} introduces the system model of temporally correlated massive access in the multi-cell network.
In Section \ref{sec:detection_strategy}, we propose the generalized sliding-window detection strategy for performance enhancement.
Then the cooperative AD and CE algorithm for the multi-cell network is introduced in Section \ref{sec:framework_design}.
Next, two cooperative AD schemes based on the proposed algorithm for the scenario with finite fronthaul capacity are investigated in Section \ref{sec:finite_capacity}.
The numerical results of the proposed algorithms are given in Section \ref{sec:simulation}.
Finally, we conclude this paper in Section \ref{sec:conclusion}.

In this paper, upper-case and lower-case letters denote random variables and their realizations, respectively.
Letters $\mathbf{x}$, $\mathbf{X}$, and $\mathcal{X}$ denote vector, matrix, and set, respectively.
Superscripts $(\cdot)^T$ and $(\cdot)^*$ denote transpose and conjugate, respectively.
Further, $\mathbb{E}[\cdot]$ and $\mathbb{V}[\cdot]$ denote expectation operation and variance operation, respectively; $|\cdot|$ denotes the magnitude of a variable or the Cardinality of a set, depending on the context.
Operator $\odot$ denotes the Hadamard product of two vectors. Finally, $\mathcal{CN}(x;\mu,\nu)$ denotes that the random variable $x$ follows complexity Gaussian distribution with mean $\mu$ and variance $\nu$.

\section{System Model}\label{sec:system_model}

\subsection{Network Architecture and Signal Model}\label{subsec:NASM}

We consider a cooperative multi-cell network consisting of $U$ cells to serve a massive number $N$ of IoT devices. We adopt the C-RAN architecture \cite{Checko_2015_CST} where all the APs are connected to the CU via fronthaul links. The fronthaul links are assumed to have infinite capacity in Section \ref{sec:framework_design}, and finite capacity in Section \ref{sec:finite_capacity}. By aggregating the content from all APs, the CU can perform joint signal processing to realize AP cooperation. The two-phase GF random access protocol is considered in this work, which comprises a pilot phase and a data phase. This work focuses on the joint AD and CE problem in the pilot phase. To facilitate the algorithm design, we assume that the APs and users are all equipped with a single antenna. We will also illustrate that the proposed algorithm can be easily extended to the case where each AP has $M \ge 2$ antennas. The main notations used in this paper are summarized in Table \ref{Tab:Notations_SM}.
\begin{table}[t]
  \centering
 \caption{Summary of Notations}\label{Tab:Notations_SM}
  \begin{tabular}{|c|c|}
    \hline
    \textbf{Notation} & \textbf{Description} \\
    \hline
    $\mathcal{N}_0$ & set of all users \\ \hline
    $U$                          & number of cells (or APs) \\ \hline
    $\mathcal{U}_n$ & AP connection set of user $n$ \\ \hline
    $\mathcal{N}_u$ & detection user set of AP $u$ \\ \hline
    $\lambda^t_n$    & activity of user $n$ in the $t$th frame \\ \hline
    $\rho_0$               & transmit power \\ \hline
    $x^t_{n,u}$           & effective channel between user $n$ and AP $u$ in the $t$th frame \\ \hline
    $g_{n,u}$               & path loss of the channel between user $n$ and AP $u$ \\ \hline
    $\mathbf{a}_n$   & pilot sequence of user $n$ \\ \hline
    $\mathbf{w}^t_{u}$ & noise vector at AP $u$ in the $t$th frame \\ \hline
    $\mathbf{z}^t_u$& noiseless received signal at AP $u$ in the $t$th frame \\ \hline
    $\alpha, \beta$ & transition probabilities \\ \hline
    $p_a$                      & active probability \\ \hline
    $(\tilde{\sigma}^{t,i}_{w,u})^2$ & estimated effective noise variance in the $i$th iteration \\ \hline
    $\mathcal{T}_w$ & sliding window \\ \hline
    $\mathcal{T}_f$   & target sub-window \\ \hline
    $\Delta_w$           & sliding step size \\ \hline
    $b^r_y$                  & quantization resolution for QF in the real number field \\ \hline
  \end{tabular}
\end{table}

Let $\mathcal{N}_0$ denote the set of all the $N$ users in the network. Each user $n \in \mathcal{N}_0$ is pre-allocated with a unique pilot sequence of length $L$, denoted as $\mathbf{a}_{n} = [a_{1,n},a_{2,n},\dots,a_{L,n}]^T \in \mathbb{C}^{L \times 1}$, for identification and channel estimation, where $L$ is assumed to be much smaller than $N$, i.e., $L \ll N$. Here, each entry in the pilot sequence is generated from the independent and identically distributed (i.i.d.) circularly symmetric complex Gaussian distribution, i.e., $\mathbf{a}_{n} \sim \mathcal{CN}(0,\frac{1}{L}\mathbf{I})$.
Since the data traffic is usually sporadic, only a small fraction of users are activated for data transmission in each frame while the others keep silent. We define a boolean variable $\lambda^t_{n} \in \{0,1\}$ to indicate the activity of each user $n$ in the $t$th frame as
\begin{align}\label{equ:lam_tn}
    \lambda^t_{n} = \left\{\begin{array}{ll}
                           1, & \text{if user $n$ is activated in the $t$th frame,} \\
                           0, & \text{otherwise}.
                         \end{array}\right.
\end{align}

Similar to \cite{Chen_2018_TSP,Liu_2018_TSP, Senel_2018_TCOM,Ke_2020_TSP,Zou_2020_SPL,Fengler_2021_TIT,Chen_2021_TIT,Xu_2015_ICC,Utkovski_2017_SPL, Chen_2019_TWC,Ke_2021_JSAC,Shao_2020_TSP,JDD_2021_TWC,Chen_2021_TSP,Wang_2021_ISIT,Zhu_2022_WCL, Zhu_2023_TCOM}, the block-fading channel model is adopted in this work, such that the channels of all users keep constant in each transmission frame but change from frame to frame. By considering Rayleigh fading channel model \cite{Chen_2019_TWC,Xu_2015_ICC,Utkovski_2017_SPL}, the channel between user $n \in \mathcal{N}_0$ and AP $u \in \{1,2,\dots,U\}$ in each frame $t$, denoted as $h^t_{n,u}$, is assumed to follow the independent complex Gaussian distribution, i.e., $h^t_{n,u} \sim \mathcal{CN}(0,g_{n,u})$, where $g_{n,u}$ is the path loss of the channel between user $n$ and AP $u$. Following \cite{Chen_2018_TSP,Liu_2018_TSP,Ke_2020_TSP,Zou_2020_SPL,Fengler_2021_TIT,Chen_2021_TIT,Xu_2015_ICC,Utkovski_2017_SPL}, we assume that the transmit powers of all users are identical and the pilot signals from the active users are perfectly synchronized at each AP. Then the superimposed pilot signals received at AP $u$ in the $t$th frame can be expressed as
\begin{align}\label{equ:Rece_APu}
    \mathbf{y}^t_u &=  \sum_{n=1}^{N} \sqrt{\rho}_0 \lambda^t_{n} h^t_{n,u} \mathbf{a}_{n} + \mathbf{w}^t_u =  \mathbf{A} \mathbf{x}^t_u + \mathbf{w}^t_u = \mathbf{z}_u^t + \mathbf{w}^t_u,
\end{align}
where $\rho_0$ is the transmit power of each user;
$\mathbf{A} = [\mathbf{a}_{1},\mathbf{a}_{2},\dots,\mathbf{a}_{N}] \in \mathbb{C}^{L \times N}$ is the pilot matrix of all users;
$\mathbf{x}^t_u \triangleq \rho_0 \pmb{\lambda}^t \odot \mathbf{h}^t_{u} = [x^t_{1,u},x^t_{2,u},\dots,x^t_{N,u}]^T \in \mathbb{C}^{N \times 1}$ represents the effective channel vector from all users to AP $u$ with $\pmb{\lambda}^{t} = [\lambda^{t}_{1},\lambda^{t}_{2},\dots,\lambda^{t}_{N}]^{T} \in \mathbb{C}^{N \times 1}$, and $\mathbf{h}^{t}_{u} = [h^{t}_{1,u},\dots,h^{t}_{N,u}]^T$; $\mathbf{z}^t_u \triangleq \mathbf{A}\mathbf{x}^t_u \in \mathbb{C}^{N \times 1}$ is defined as the noiseless received signal. We also denote the received signal matrix and effective channel matrix over all $U$ cells in frame $t$ as $\mathbf{Y}^t = [\mathbf{y}^t_1,\dots,\mathbf{y}^t_U]$ and $\mathbf{X}^t = [\mathbf{x}^t_1,\dots,\mathbf{x}^t_U]$, respectively; $\mathbf{w}^t_u = [w^t_{1,u},w^t_{2,u},\dots,w^t_{l,u}] \in \mathbb{C}^{L \times 1}$ is the i.i.d. additive white Gaussian noise vector at AP $u$ in the $t$th frame with zero mean and variance $\sigma^2_w$.
The variance $\sigma^2_w$ can be known by parameter estimation schemes \cite{Das_2012_TSP}, e.g., maximum likelihood estimation.
By collecting the received signals from all APs, the CU can perform joint AD and CE.

\subsection{Evolution Process of User Activity}\label{sec:sub_AEP}

Following \cite{Jiang_2021_TWC,Wang_2021_ISIT,Zhu_2023_TCOM,Zhu_2022_WCL}, we adopt a first-order Markov process to model the temporal correlation of user activity. Specifically, we assume that the activity evolution of each user follows an i.i.d. first-order steady Markov process with two discrete states\footnote{Our proposed methods can be easily modified to handle non-identically distributed user activities.}.
The Markov process for each user $n$ can be characterized by four transition probabilities $\text{Pr}(\lambda^t_{n}|\lambda^{t-1}_{n})$s.
For simplicity, we denote $\text{Pr}(\lambda^t_{n} = 1|\lambda^{t-1}_{n} = 0) = \alpha$ and $\text{Pr}(\lambda^t_{n} = 1|\lambda^{t-1}_{n} = 1) = \beta$ for all $t,n$. Then, the other two transition probabilities can be easily obtained as $\text{Pr}(\lambda^t_{n} = 0|\lambda^{t-1}_{n} = 0) = 1 - \alpha$ and $\text{Pr}(\lambda^t_{n} = 0|\lambda^{t-1}_{n} = 1) = 1 - \beta$.
In the special case when $\alpha=\beta$, we have $\text{Pr}(\lambda^t_{n}|\lambda^{t-1}_{n}) = \text{Pr}(\lambda^t_{n})$, which means that there is no temporal correlation in the activity of user $n$. The larger $\beta$ is, the stronger temporal correlation of user $n$ has.

Under the steady-state condition, the active probability of each user $n$ in each frame $t$ is a constant and denoted as $\textbf{Pr}(\lambda^t_{n} = 1) = p_a$. By the law of total probability, we have $\alpha (1-p_a)+\beta p_a = p_a$ and thus obtain $p_a = \alpha/(1+\alpha-\beta)$.
Based on such statistical relationship among $\alpha$, $\beta$, and $p_a$, we can utilize the probabilities $\beta$ and $p_a$ to fully describe the activity evolution process in the rest of this paper.
The exact values of $\beta$ and $p_a$ are assumed to be known priorly in this work, which can also be accurately estimated using machine learning techniques \cite{Zhu_2023_TCOM, Zhu_2021_TWC}.

\subsection{User-Centric AP Cooperation}

\begin{figure}[t]
  \centering
  \includegraphics[width=.4\textwidth]{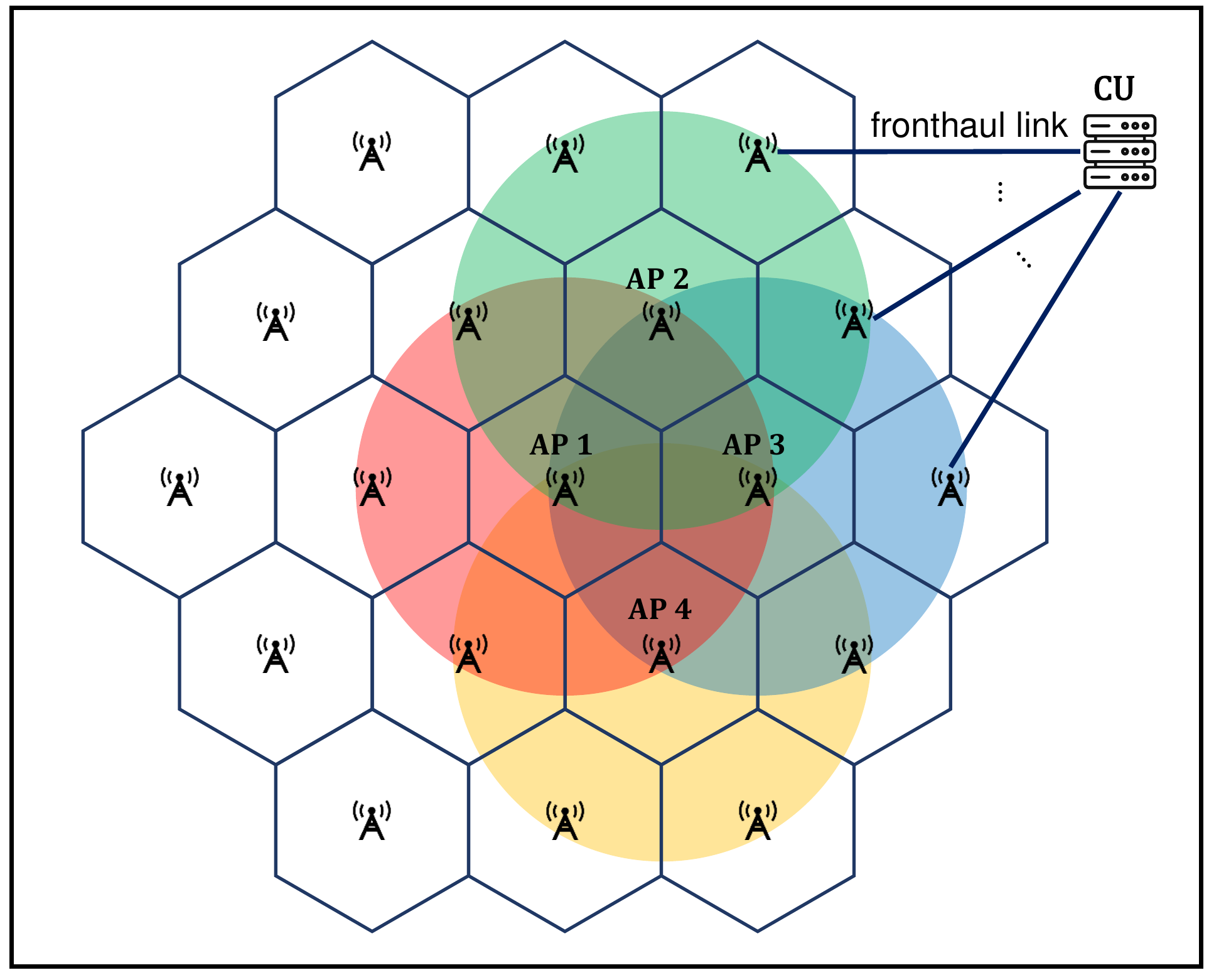}
  \caption{The diagram of the cooperative multi-cell network where each AP detects the users in a circle area with radius $D_{max}$ centered on the AP. For example, AP $1$ performs joint AD and CE on the users in the red circle.}\label{Fig:C-RAN_Model}
\end{figure}

In the multi-cell network, it is impractical to involve all the APs to detect the activity of every user in the network due to prohibitively high complexity. It is also unnecessary to do so since for those APs far from a target user, the signals received by these APs for the activity detection of this user are unreliable.
Therefore, we propose the user-centric AP cooperation strategy to decide which subset of APs to collaboratively perform AD and CE on user $n$.
By assuming that all users are uniformly distributed at random in the network \cite{Chen_2019_TWC}, we define the AP connection set of each user $n$ as
\begin{align}\label{equ:Set_Un}
    \mathcal{U}_{n} &= \{ u | d_{n,u} \le D_{max}, u \in \{1,2,\dots,U\} \},
\end{align}
where $d_{n,u}$ is the distance between user $n$ and AP $u$ and $D_{max}$ is a pre-defined maximal detection distance. Accordingly, the detection user set $\mathcal{N}_u$ for each AP $u$ can be defined as
\begin{align}\label{equ:Set_Nu}
    \mathcal{N}_u &= \{ n | d_{n,u} \le D_{max}, n \in \mathcal{N}_0 \}.
\end{align}

Here, the maximal detection distance is determined by empirical results.
From the definition in (\ref{equ:Set_Un}) and (\ref{equ:Set_Nu}), the sets of $\mathcal{U}_{n}$ and $\mathcal{N}_{u}$ are in fact derived based on the statistical channel information of the path loss $g_{n,u}$. The diagram of the multi-cell network with user-centric AP cooperation is shown in Fig. \ref{Fig:C-RAN_Model}.
Nevertheless, the sets of $\mathcal{U}_{n}$ and $\mathcal{N}_{u}$ with other definitions can also be employed for the specific systems.

Accounting the temporally correlated user activity, we consider the joint AD and CE problem in $T \ge 2$ consecutive frames.
This is equivalent to solving the linear inverse problem that recovers the set of effective channel matrices $\{\mathbf{X}^t\}_{t=1}^{T}$ from the set of received signal matrices $\{\mathbf{Y}^t\}_{t=1}^{T}$ given the pilot matrix $\mathbf{A}$.
Note that the joint AD and CE problem for multi-cell networks is usually formulated as a CS-MMV problem in \cite{Xu_2015_ICC, Chen_2019_TWC,Ke_2021_JSAC}, which however considers only one frame at a time without taking the temporal correlation of user activity into account.
In this paper, we shall solve the joint AD and CE problem from the DCS-MMV perspective, where the received signals in multiple consecutive frames are jointly processed. However, the existing optimization-based methods and greedy-based algorithms for the DCS-MMV problem usually have very high computational complexity in large-scale problems and ignore the utilization of system statistics. Thus, we resort to the Bayesian inference method to provide the optimal mean square error (MSE) estimation.

\section{Generalized Sliding-Window Detection Strategy}\label{sec:detection_strategy}

\begin{figure}[t]
  \centering
  \includegraphics[width=.45\textwidth]{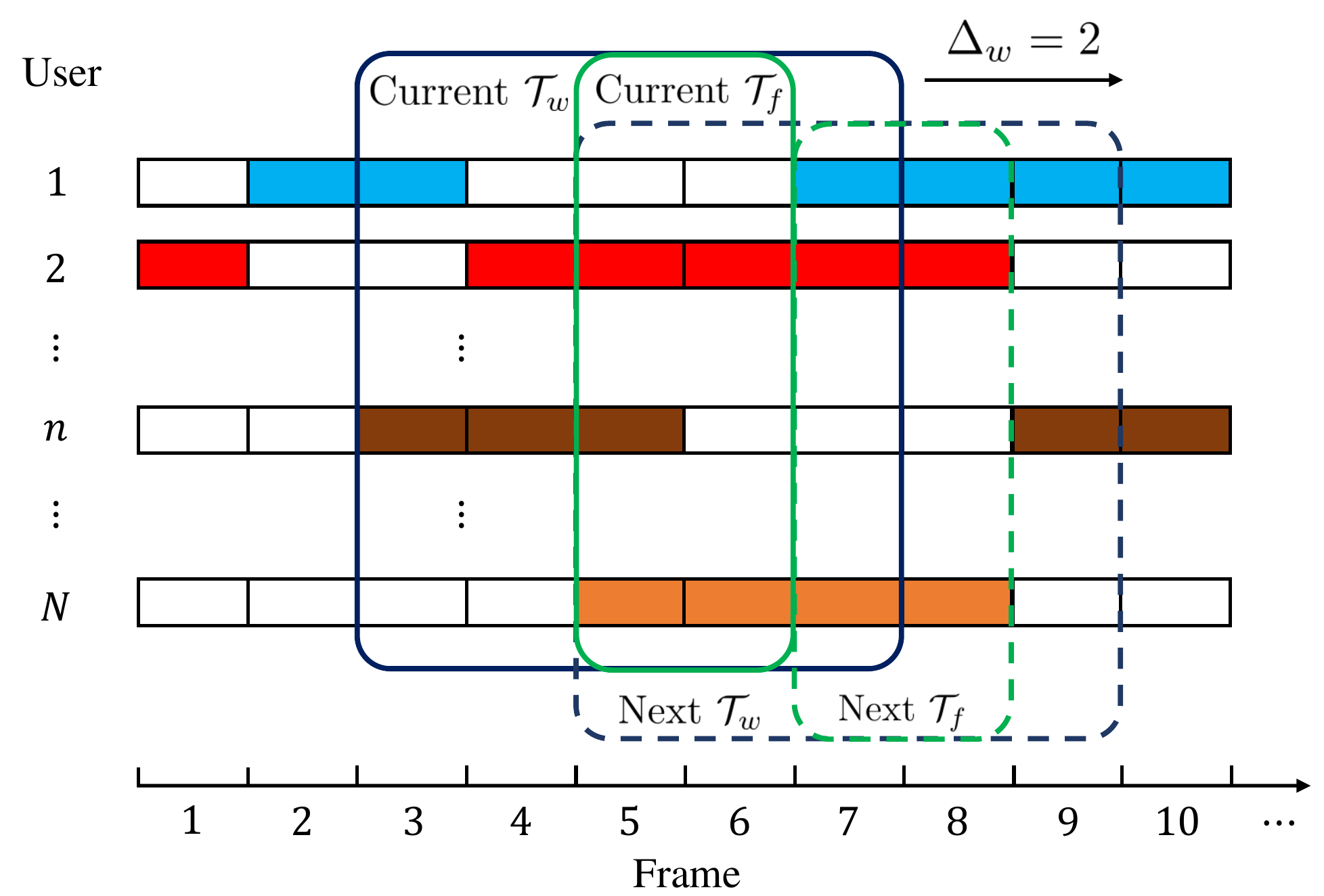}
  \caption{An example of the generalized sliding-window detection strategy with $T_w = 5,\Delta_w = 2$. (The solid rectangle means user $n$ being active in this frame, whereas the blank rectangle means user $n$ being inactive in this frame.)}\label{Fig:SWS}
\end{figure}

The proposed AD and CE algorithm in this paper is built upon a generalized sliding-window detection strategy. This strategy  facilitates the full utilization of the temporally correlated user activity and includes all the existing sliding-window detection strategies in \cite{Jiang_2021_TWC,Wang_2021_ISIT,Zhu_2023_TCOM,Zhu_2022_WCL} as special cases. In this section, we present the generalized sliding window detection strategy in detail.

Let the sliding window be denoted as $\mathcal{T}_w = \{t_0,\dots,t_0+T_w-1\}$. It is specifically characterized by three parameters: window size $T_w = |\mathcal{T}_w|$, target sub-window $\mathcal{T}_f =\{t_1,\dots,t_1+\Delta_w-1\} \subseteq \mathcal{T}_w$ with $t_0 \le t_1 \le t_0+T_w-\Delta_w$, and sliding step size $\Delta_w = |\mathcal{T}_f| \le T_w$.
Note that the user activities over the target sub-window $T_f$ shall be detected at once together.
Thus, our proposed detection strategy is a block-by-block detection rather than a frame-by-frame detection strategy.
An example with $T_w = 5$ and $\Delta_w = 2$ is also given in Fig. \ref{Fig:SWS}. In the current sliding window $\mathcal{T}_w = \{3,\dots,7\}$, we can set $\mathcal{T}_f = \{ 5,6\}$. After the sliding window moves to the next position with $\mathcal{T}_w = \{5,\dots,9\}$, the target sub-window will be accordingly changed to $\mathcal{T}_f = \{ 7,8\}$.
In fact, we can have various choices of $\mathcal{T}_f$. By the block-by-block detection, the average detection latency is $\bar{\tau}_D = t_0+T_w-t_1 - \frac{\Delta_w+1}{2}$.

In the special case where $\mathcal{T}_w = \{t_0, t_0+1\}$ with $\Delta_w = 1$ and $\mathcal{T}_f = \{t_0+1\}$, the above generalized sliding-window detection reduces to the schemes in \cite{Jiang_2021_TWC, Wang_2021_ISIT}.
These works \cite{Jiang_2021_TWC, Wang_2021_ISIT} can only exploit the historical estimations for frame-by-frame detection. In the special case where $\mathcal{T}_f = \mathcal{T}_w$ with $\Delta_w \ge 2$, it reduces to the schemes in \cite{Zhu_2023_TCOM}. However, this scheme is a pure block-by-block detection without exploiting the temporal correlation across different blocks.

\section{Joint AD and CE Algorithm}\label{sec:framework_design}

In this section, we employ the Bayesian inference method to solve the DCS-MMV problem based on the generalized sliding-window detection strategy introduced in the previous section. By building the factor graph based on the probabilistic model, we develop a scalable DCS-MMV-GAMP algorithm to realize joint AD and CE in multi-cell massive access.

\subsection{Bayesian Inference and Graphical Representation}

\begin{figure}[t]
  \centering
  \includegraphics[width=0.5\textwidth]{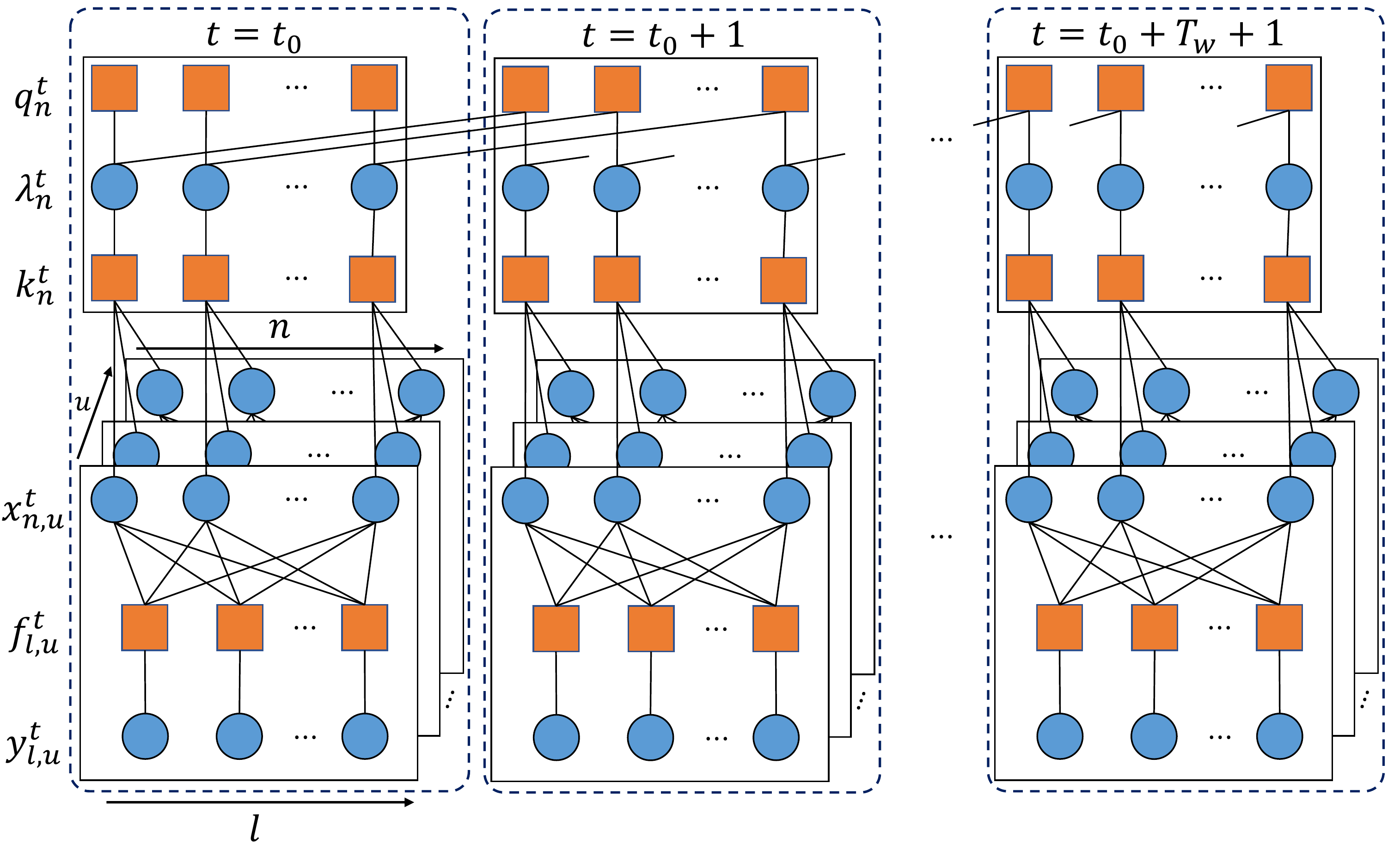}
  \caption{Factor graph of the joint posterior distribution (\ref{equ:PP_xlam}) for the received signals in the sliding window $\mathcal{T}_w$.}\label{Fig:FG}
\end{figure}

With system statistics available, the Bayesian inference method can be applied to obtain the optimal solution of our problem. First, we give the prior distribution of the effective channel vector $\mathbf{x}^t_{n} = [x^t_{n,1},x^t_{n,2},\dots,x^t_{n,U}] \in \mathbb{C}^{1 \times U}$ from each user $n$ to all the APs in the $t$th frame as a multi-variable Bernoulli-Gaussian distribution
\begin{align}\label{equ:P_x}
    p(\mathbf{x}^t_{n}|\lambda^t_{n}) &= (1-\lambda^{t}_{n}) \cdot \delta(\mathbf{x}_{n}) + \lambda^t_{n} \cdot \mathcal{CN}(0,\pmb{\Sigma}_{n}),
\end{align}
where $\delta(\cdot)$ is the point mass measure at $\mathbf{0}$ and $\pmb{\Sigma}_{n} = \text{diag}(\tilde{\mathbf{g}}_{n})$ is the covariance matrix with $\tilde{\mathbf{g}}_{n} = \rho_0 \mathbf{g}_{n} = \rho_0\cdot[g_{n,1},g_{n,2},\dots,g_{n,U}] \in \mathbb{C}^{1 \times U}$ representing the effective large-scale attenuation vector. The probability of the user activity $\text{Pr}(\lambda^t_{n})$ is related to $\text{Pr}(\lambda^{t-1}_{n})$ with the Markov chain as discussed in Section \ref{sec:system_model}.

Based on the received signal matrices $\{\mathbf{Y}^t\}_{t\in\mathcal{T}_w}$ over the detection window, the joint posterior probability of $\{\pmb{\lambda}^t\}_{t\in\mathcal{T}_w}$ and $\{\mathbf{X}^t\}_{t\in\mathcal{T}_w}$ can be calculated from the Bayes' rule as
\begin{align}\label{equ:PP_xlam}
    & p(\{\pmb{\lambda}^{t}\}_{t\in\mathcal{T}_w}, \{\mathbf{X}^{t}\}_{t\in\mathcal{T}_w} | \{\mathbf{Y}^{t}\}_{t\in\mathcal{T}_w}) \notag \\
     = &  \frac{1}{p(\{\mathbf{Y}^{t}\}_{t\in\mathcal{T}_w})} \Bigg( \prod_{t\in\mathcal{T}_w} \Big( p(\mathbf{Y}^{t}|\mathbf{X}^{t}) \prod_{n=1}^{N} p(\mathbf{x}^{t}_{n}|\lambda^{t}_{n})p(\lambda^{t}_{n}|\lambda^{t-1}_{n}) \Big) \Bigg),
\end{align}
where the marginal probability of the received signals $\{\mathbf{Y}\}_{t\in\mathcal{T}_w}$ can be calculated as
\begin{align}\label{equ:MP_Y}
    p(\{\mathbf{Y}^{t}\}_{t\in\mathcal{T}_w}) = &~ \int_{\{\pmb{\lambda}^{t},\mathbf{X}^{t}\}_{t\in\mathcal{T}_w}}  \prod_{t\in\mathcal{T}_w} \Big( p(\mathbf{Y}^{t}|\mathbf{X}^{t}) \notag \\
     \quad &\times \prod_{n=1}^{N} p(\mathbf{x}^{t}_{n}|\lambda^{t}_{n})p(\lambda^{t}_{n}|\lambda^{t-1}_{n}) \Big). 
\end{align}
In both (\ref{equ:PP_xlam}) and (\ref{equ:MP_Y}), at the first frame to the sliding window, we have $p(\lambda^{t_0}_{n}|\lambda^{t_0-1}_{n}) = p_a\lambda^{t_0}_{n} + (1-p_a)(1-\lambda^{t_0}_{n})$.

Under Bayesian inference, the active probability of each user $n$ in the target frames can be estimated by $\{\text{Pr}(\lambda^t_{n}|\{\mathbf{Y}^{\tau}\}_{\tau\in\mathcal{T}_w})\}_{t\in\mathcal{T}_f}$, and the corresponding effective channel coefficient is obtained as $\{\mathbb{E}[x^t_{n,u}|\{\mathbf{Y}^{\tau}\}_{\tau\in\mathcal{T}_w}]\}_{t\in\mathcal{T}_f}$ by following the minimum MSE (MMSE) criterion. However, the calculations of the posterior probabilities $\text{Pr}(\lambda^t_{n}|\{\mathbf{Y}^{\tau}\}_{\tau\in\mathcal{T}_w})$ and $\text{Pr}(x^t_{n,u}|\{\mathbf{Y}^{\tau}\}_{\tau\in\mathcal{T}_w})$ are usually intractable due to the very large dimensional integrals. In the next subsection, we introduce a scalable algorithm that can approximately achieve the Bayes-optimal performance.

\begin{table}
  \centering
  \caption{The Underlying Distributions of the Factor in the Factor Graph.}\label{Tab:UD_FG}
  \begin{tabular}{c|ccc}
    \hline
    \hline
    \textbf{Factor} & $f^t_{l,u}$ & $k^t_{n}$ & $q^t_{n}$ \\
    \hline
    \textbf{Distribution} & $p(y^t_{l,u}|\mathbf{x}^t_{u})$ & $p(\mathbf{x}^t_{n}|\lambda^t_{n})$ &$p(\lambda^t_{n}|\lambda^{t-1}_{n})$ \\
    \hline
    \hline
  \end{tabular}
\end{table}

\begin{table}[t]
  \centering
  \caption{The Notation of the Updated Probability for $\lambda^t_{n} = 1$ in the Extrinsic Message at the $i$th Iteration.}\label{Tab:NUP}
  \begin{tabular}{cc}
    \hline
    \hline
    \textbf{Extrinsic Message} & \textbf{Probability} \\
    \hline
    $\xi^{t,i}_{x^t_{n,u} \rightarrow k^t_{n}}(\lambda^t_{n})$ & $\overleftarrow{\phi}^{t,i}_{n,u}$ \\
    $\xi^{t,i}_{k^t_{n} \rightarrow x^t_{n,u}}(\lambda^t_{n})$ & $\overrightarrow{\phi}^{t,i}_{n,u}$ \\
    $\xi^{t,i}_{k^t_{n} \rightarrow \lambda^t_{n}}(\lambda^t_{n})$ & $\overleftarrow{\pi}^{t,i}_{n}$ \\
    $\xi^{t,i}_{\lambda^t_{n} \rightarrow k^t_{n}}(\lambda^t_{n})$ & $\overrightarrow{\pi}^{t,i}_{n}$ \\
    $\xi^{t,i}_{\lambda^t_{n} \rightarrow q^t_{n}}(\lambda^t_{n})$ & $\overleftarrow{\psi}^{t,i}_{n}$ \\
    $\xi^{t,i}_{q^t_{n} \rightarrow \lambda^t_{n}}(\lambda^t_{n})$ & $\overrightarrow{\psi}^{t,i}_{n}$ \\
    $\xi^{t,i}_{\lambda^{t}_{n} \rightarrow q^{t+1}_{n}}(\lambda^t_{n})$ & $\overleftarrow{\varphi}^{t,i}_{n}$ \\
    $\xi^{t,i}_{q^{t+1}_{n} \rightarrow \lambda^{t}_{n}}(\lambda^t_{n})$ & $\overrightarrow{\varphi}^{t,i}_{n}$ \\
    \hline
    \hline
  \end{tabular}
\end{table}

Based on the decompositions of the joint posterior probability in (\ref{equ:PP_xlam}), we establish the factor graph as shown in Fig. \ref{Fig:FG}, where an orange square represents a factor corresponding to the sub-constraint function $p(y^t_{l,u}|\mathbf{x}^t_u)$, $p(\mathbf{x}^t_{n}|\lambda^t_u)$ or $p(\lambda^t_n|\lambda^{t-1}_{n})$, whereas a blue circle represents a variable node associated with the random variable $y^t_{l,u}$, $x^t_{n,u}$ or $\lambda^t_{n}$. Table \ref{Tab:UD_FG} shows the represented distributions of the factors in the factor graph. The factor graph shows that the variable nodes are only associated with the sub-constraint functions, implying that the MP-based algorithms \cite{Ksch_2001_TIT} can be employed to simplify the calculation of the posterior probabilities $\text{Pr}(\lambda^t_{n}|\{\mathbf{Y}^{\tau}\}_{\tau\in\mathcal{T}_w})$ and $\text{Pr}(x^t_{n,u}|\{\mathbf{Y}^{\tau}\}_{\tau\in\mathcal{T}_w})$.
In general, the MP-based algorithms perform iterative extrinsic message propagations between the adjacent nodes and then calculate the \emph{posterior} probabilities of each variable by combining the messages from all its connected factors after convergence.
Thus in this paper, we use $\xi^i_{c \rightarrow e}(x)$ to denote the extrinsic message from an arbitrary node $c$ to its adjacent node $e$ on random variable $x$ in the $i$th iteration.

\subsection{DCS-Based Multiple Measurement Vector GAMP Algorithm}

To reduce the computational complexity and hence make the MP-based algorithm scalable in massive access, we combine GAMP and MP together, inspired by the HyGAMP framework \cite{Rangan_2017_TSP}. In specific, the GAMP approximations are utilized to simplify the MP update between the nodes $x^t_{n,u}$ and $f^t_{l,u}$, while the other messages in the factor graph are still updated by the standard MP principle.
Due to the user-centric AP cooperation, the factor graph is simplified, where each node $x^t_{n,u}$ is only connected to the factors $k^t_{n}$ with $n \in \mathcal{N}_u$ and the factors $f^t_{l,u}$ with $u \in \mathcal{U}_{n}$.
Note that for the generalized sliding-window detection strategy, the final results comprise of only the estimation of $\{\text{Pr}(\lambda^t_{n}|\{\mathbf{Y}^{\tau}\}_{\tau\in\mathcal{T}_w})\}_{t\in\mathcal{T}_w}$ and $\{\mathbb{E}[x^t_{n,u}|\{\mathbf{Y}^{\tau}\}_{\tau\in\mathcal{T}_w}]\}_{t\in\mathcal{T}_w}$, which can exploit the side information from both their neighboring frames. Accordingly, we develop a DCS-MMV-GAMP algorithm to further simplify the general HyGAMP framework \cite{Rangan_2017_TSP} and
to boost the estimation performance over the conventional methods \cite{Zou_2020_SPL, Jiang_2021_TWC, Wang_2021_ISIT}.

We can divide the DCS-MMV-GAMP algorithm into the activity refinement part and the CE part. In each iteration, the activities are first refined based on their statistical dependencies in the spatial-temporal domain via standard MP. Then the channel coefficient $x^t_{n,u}$ with $n \in \mathcal{N}_u$ is recovered from the received signal $\mathbf{y}^{t}_{u}$ via GAMP and we force the estimation $\widehat{x}^t_{n,u} = 0$ for all $n \notin \mathcal{N}_u$. Note that these two parts also exchange extrinsic messages in each iteration, which enhances the performance of both AD and CE. We outline the proposed algorithm in Algorithm \ref{Alg:DCS-MMV-GAMP} and give the detailed derivation of DCS-MMV-GAMP in the following.


\subsubsection{Activity Refinement}
The activity refinement part aims to perform the initialization and updates the active probabilities based on the CE results in the last iteration. This part includes factors $\{q^{t}_{n,u}\}_{t=t_0,n=1}^{t_0+T_w-1,N}$, $\{k^{t}_{n}\}_{t=t_0,n=1}^{t_0+T_w-1,N}$ and variables $\{\lambda^t_{n}\}_{t=t_0,n=1}^{t_0+T_w-1,N}$. For clarification, the notations of the active probabilities updated in the extrinsic messages are given in Table \ref{Tab:NUP}.

In the $i$th iteration, the message conveyed from $k^{t}_{n}$ to $\lambda^{t}_{n}$ is first obtained by combining the extrinsic messages propagated from the associated $f^t_{l,u}$s with $u \in \mathcal{U}_{n}$, which can be expressed by
\begin{align}\label{equ:M_k_to_lam}
    \xi^{i}_{k^{t}_{n} \rightarrow \lambda^{t}_{n}}(\lambda^t_{n})  \propto&~ \prod_{u \in \mathcal{U}_{n}} \xi^i_{x^t_{n,u} \rightarrow k^t_{n}}(\lambda^t_{n}),
\end{align}
where the active probability $\overleftarrow{\pi}^{t,i}_{n}$ is updated by
\begin{align}\label{equ:leftarrow_pi}
    \overleftarrow{\pi}^{t,i}_{n} &= \frac{\prod_{u \in \mathcal{U}_{n}} \overleftarrow{\phi}^{t,i}_{n,u}}{\prod_{u \in \mathcal{U}_{n}}(1-\overleftarrow{\phi}^{t,i}_{n,u}) + \prod_{u \in \mathcal{U}_{n}} \overleftarrow{\phi}^{t,i}_{n,u}}.
\end{align}
To fully exploit the temporal correlation between the activities in the adjacent frames, the proposed algorithm performs both forward and backward message propagations across the frames.
The forward messages between the adjacent frames can be calculated as
\begin{align}
    \xi^{i}_{\lambda^{t}_{n} \rightarrow q^{t+1}_{n}}(\lambda^{t}_{n}) &\propto \xi^{i}_{k^{t}_{n} \rightarrow \lambda^{t}_{n}}(\lambda^t_{n}) \cdot \xi^{i}_{q^{t}_{n} \rightarrow \lambda^{t}_{n}}(\lambda^t_{n}), \label{equ:lamt_to_qt+1} \\
    \xi^{i}_{q^{t}_{n} \rightarrow \lambda^{t}_{n}}(\lambda^{t}_{n}) &\propto \int_{\lambda^{t-1}_{n}} p(\lambda^{t}_{n}|\lambda^{t-1}_{n}) \cdot \xi^{i}_{\lambda^{t-1}_{n} \rightarrow q^{t}_{n}}(\lambda^{t-1}_{n}) , \label{equ:qt_to_lamt}
\end{align}
and the backward messages can be obtained as
\begin{align}
    \xi^{i}_{\lambda^{t}_{n} \rightarrow q^{t}_{n}}(\lambda^{t}_{n}) &\propto \xi^{i}_{k^{t}_{n} \rightarrow \lambda^{t}_{n}}(\lambda^t_{n}) \cdot \xi^{i}_{q^{t+1}_{n} \rightarrow \lambda^{t}_{n}}(\lambda^t_{n}), \label{equ:lamt_to_qt} \\
    \xi^{i}_{q^{t+1}_{n} \rightarrow \lambda^{t}_{n}}(\lambda^{t}_{n}) &\propto \int_{\lambda^{t+1}_{n}} p(\lambda^{t+1}_{n}|\lambda^{t}_{n}) \cdot \xi^{i}_{\lambda^{t+1}_{n} \rightarrow q^{t+1}_{n}}(\lambda^{t+1}_{n}). \label{equ:qt+1_to_lamt}
\end{align}
Therein, the updated probabilities in these messages are obtained as
\begin{align}
    \overleftarrow{\varphi}^{t,i}_{n} &= \frac{\overleftarrow{\pi}^{t,i}_{n} \overrightarrow{\psi}^{t,i}_{n}}{(1-\overleftarrow{\pi}^{t,i}_{n})(1-\overrightarrow{\psi}^{t,i}_{n}) + \overleftarrow{\pi}^{t,i}_{n} \overrightarrow{\psi}^{t,i}_{n}}, \label{equ:leftarrow_varphi}  \\
    \overrightarrow{\psi}^{t,i}_{n} &= \alpha (1-\overleftarrow{\varphi}^{t-1,i}_{n}) + \beta \overleftarrow{\varphi}^{t-1,i}_{n}, \label{equ:rightarrow_psi} \\
    \overleftarrow{\psi}^{t,i}_{n} &= \frac{\overleftarrow{\pi}^{t,i}_{n} \overrightarrow{\varphi}^{t,i}_{n}}{(1-\overleftarrow{\pi}^{t,i}_{n}) (1-\overrightarrow{\varphi}^{t,i}_{n})+\overleftarrow{\pi}^{t,i}_{n} \overrightarrow{\varphi}^{t,i}_{n}},  \label{equ:leftarrow_psi} \\
    \overrightarrow{\varphi}^{t,i}_{n} &= \frac{(1-\beta)(1-\overleftarrow{\psi}^{t+1,i}_{n}) + \beta \overleftarrow{\psi}^{t+1,i}_{n}}{(2-\alpha-\beta)(1-\overleftarrow{\psi}^{t+1,i}_{n}) +(\alpha+\beta)\overleftarrow{\psi}^{t+1,i}_{n}}. \label{equ:rightarrow_varphi}
\end{align}
For $t = t_0$, we always have $\xi^{i}_{q^{t}_{n} \rightarrow \lambda^{t}_{n}}(\lambda^{t}_{n}) = (1-\lambda^{t}_{n})(1-p_a) + \lambda^{t}_{n}p_a$ and thus $\overrightarrow{\psi}^{t_0,i}_{n} = p_a$. For $t = t_0 + T_w - 1$, we have $\overleftarrow{\varphi}^{t,i}_{n} = \overleftarrow{\pi}^{t,i}_{n}$ by $\xi^{i}_{\lambda^{t}_{n} \rightarrow q^{t}_{n}}(\lambda^{t}_{n}) = \xi^{i}_{k^{t}_{n} \rightarrow \lambda^{t}_{n}}(\lambda^{t}_{n})$.
If the distributions of the user activities are not identical, we can replace $\{\alpha, \beta\}$ by $\{\alpha_n, \beta_n\}$ in equations (\ref{equ:rightarrow_psi}) and (\ref{equ:rightarrow_varphi}) and set $\overrightarrow{\psi}^{t_0,i}_{n} = p_n$ for each user $n$, where $\{\alpha_n, \beta_n, p_n\}$ is the statistical parameter set of the evolution process for user $n$.
Note that the above probabilities (\ref{equ:leftarrow_varphi})-(\ref{equ:rightarrow_varphi}) are not updated simultaneously. In particular, the probabilities $\overrightarrow{\psi}^{t,i}_{n}$ and $\overleftarrow{\varphi}^{t,i}_{n}$ are computed sequentially from the $t_0$th frame to the $(t_0+T_w-1)$th frame,
while the other two probabilities $\overleftarrow{\psi}^{t,i}_{n}$ and $\overrightarrow{\varphi}^{t,i}_{n}$ are conversely updated from the $(t_0+T_w-1)$th frame to the $t_0$th frame.
After these probabilities passed serially or conversely through these $T_w$ frames, the message $\xi^{i}_{\lambda^{t}_{n} \rightarrow k^{t}_{n}}(\lambda^{t}_{n})$ can be given as
\begin{align}\label{equ:M_lam_to_k}
    \xi^{i}_{\lambda^{t}_{n} \rightarrow k^{t}_{n}}(\lambda^{t}_{n}) &\propto \xi^{i}_{q^{t}_{n} \rightarrow \lambda^{t}_{n}}(\lambda^{t}_{n}) \cdot \xi^{i}_{q^{t+1}_{n} \rightarrow \lambda^{t}_{n}}(\lambda^{t}_{n}).
\end{align}
Thus, the active probability updated in $\xi^{i}_{\lambda^{t}_{n} \rightarrow k^{t}_{n}}(\lambda^{t}_{n})$ can be obtained as
\begin{align}\label{equ:P_lam_to_k}
    \overrightarrow{\pi}^{t,i}_{n} &= \frac{\overrightarrow{\psi}^{t,i}_{n}\overrightarrow{\varphi}^{t,i}_{n}}{(1-\overrightarrow{\psi}^{t,i}_{n})(1-\overrightarrow{\varphi}^{t,i}_{n}) + \overrightarrow{\psi}^{t,i}_{n}\overrightarrow{\varphi}^{t,i}_{n}}.
\end{align}
After that, the message conveyed to refine the effective channel coefficient between user $n$ with its connected AP $u$ at each frame $t$ can be expressed as
\begin{align}\label{equ:M_k_to_x}
    \xi^{i}_{k^{t}_{n} \rightarrow x^{t}_{n,u}}(\lambda^{t}_{n}) &\propto \xi^{i}_{\lambda^{t}_{n} \rightarrow k^{t}_{n}}(\lambda^{t}_{n}) \prod_{\substack{v \in \mathcal{U}_{n} \\ v \ne u}} \xi^{i}_{x^{t}_{n,v} \rightarrow k^{t}_{n}}(\lambda^{t}_{n}), \forall u \in \mathcal{U}_{n}.
\end{align}
The refined probability in $\xi^{i}_{k^{t}_{n} \rightarrow x^{t}_{n,u}}(\lambda^{t}_{n})$ is therefore given by
\begin{align}\label{equ:P_k_to_x}
    \overrightarrow{\phi}^{i}_{n,u} &= \frac{\overrightarrow{\pi}^{t,i}_{n}\prod_{\substack{v \in \mathcal{U}_{n},\\ v \ne u}}
    \overleftarrow{\phi}^{t,i}_{n,v}}{(1-\overrightarrow{\pi}^{t,i}_{n})\prod_{\substack{v \in \mathcal{U}_{n},\\ v \ne u}}(1-\overleftarrow{\phi}^{t,i}_{n,v}) + \overrightarrow{\pi}^{t,i}_{n}\prod_{\substack{v \in \mathcal{U}_{n},\\ v \ne u}}\overleftarrow{\phi}^{t,i}_{n,v}}.
\end{align}

\subsubsection{Channel Estimation}

After refining the activity detection, this part performs MMSE estimation on $x^t_{n,u}$ with the updated extrinsic message $\xi^{i}_{k^{t}_{n} \rightarrow x^{t}_{n,u}}(\lambda^{t}_{n})$. Based on the factor graph, the messages exchanged between the factor node $x^{t}_{n,u}$ and the variable node $f^{t}_{l,u}$ from the standard MP principle can be expressed as
\begin{align}
    \xi^{i}_{x^{t}_{n,u} \rightarrow f^{t}_{l,u}}(x^t_{n,u}) &\propto \xi^{i}_{k^{t}_{n} \rightarrow x^{t}_{n,u}} \prod_{s \ne l} \xi^{i}_{f^{t}_{s,u} \rightarrow x^{t}_{n,u}}(x^t_{n,u}),~ \forall n \in \mathcal{N}_{u}, \label{equ:x_to_f} \\
    \xi^{i}_{f^{t}_{l,u} \rightarrow x^{t}_{n,u}}(x^t_{n,u}) &\propto \int_{\mathbf{x}^{t}_{u} \setminus x^{t}_{n,u}} p(y^{t}_{l,u}|\mathbf{x}^{t}_{u})\prod_{\substack{o \in \mathcal{N}_{u}, \\ o \ne n}} \xi^{i}_{x^{t}_{o,u} \rightarrow f^{t}_{l,u}}(x^t_{n,u}). \label{equ:f_to_x}
\end{align}
However, the high-dimensional integrals in (\ref{equ:f_to_x}) still hinder the standard MP algorithm to be implemented in large-scale systems. As a solution, the GAMP approximation is leveraged to simplify the messages (\ref{equ:x_to_f}) and (\ref{equ:f_to_x}) with Gaussian distributions in the large-scale system limit \cite{Rangan_2017_TSP}, which can shrink the number of update variables from $\mathcal{O}(T_wLNU)$ to $\mathcal{O}(T_wU(L+N))$. In particular, for the linear inverse problem that recovers $\mathbf{x}_{u}$ from $\mathbf{y}_{u}$ with given $\mathbf{A}$, the GAMP algorithm can handle arbitrary prior input distribution $p(\mathbf{x}^{t}_{u})$ and output distribution $p(\mathbf{y}^{t}_{u}|\mathbf{A}\mathbf{x}^{t}_{u})$.

Now we illustrate the GAMP method \cite{Rangan_2011_ISIT} used in Algorithm \ref{Alg:DCS-MMV-GAMP}. First, the variable $\widehat{p}^{t,i}_{l,u}$ associated with the variance $\nu^{z,t,i}_{l,u}$ is calculated in Lines 8-9, which is defined as the plug-in estimate of the noiseless signal $z^{t}_{l,u}$. Under the MMSE criterion, the posterior mean and variance of $z^{t}_{l,u}$ in the $i$th iteration can be estimated in Lines 10-11. Subsequently, the residual $\widehat{s}^{t,i}_{l,u}$ and its variance $\nu^{s,t,i}_{l,u}$ are obtained in Lines 12-13. In Lines 14-15, the variables $\widehat{r}^{t,i}_{n,u}$ and $\nu^{r,t,i}_{n,u}$ are calculated with the updated residual $\widehat{s}^{t,i}_{l,u}$, where $\widehat{r}^{t,i}_{n,u}$ is statistically equal to the true channel coefficient $x^{0,t}_{n,u}$ plus a Gaussian noise with zero mean and variance $\nu^{r,t,i}_{n,u}$ in general. With the noise-corrupted observation $\widehat{r}^{t,i}_{n,u}$ and the refined probability $\overrightarrow{\phi}^{t,i}_{n,u}$, the MMSE estimation of the effective channel coefficient with its variance is derived in Lines 16-17.

For the estimate of the noiseless signal $z^{t,i}_{l,u}$, we can obtain its posterior probability as
\begin{align}\label{equ:Pp_z}
    p(z^{t}_{l,u}|\{\mathbf{Y}^{\tau}\}_{\tau \in \mathcal{T}_w}) &= \frac{p(y^{t}_{l,u}|z^{t}_{l,u})p(z^{t}_{l,u})}{\int p(y^{t}_{l,u}|z^{t}_{l,u})p(z^{t}_{l,u}) dz^{t}_{l,u}}, 
\end{align}
where $p(z^{t}_{l,u}) = \mathcal{CN}(z^{t}_{l,u};\widehat{p}^{t,i}_{l,u},\nu^{p,t,i}_{l,u})$ and $p(y^{t}_{l,u}|z^{t}_{l,u}) = \mathcal{CN}(y^{t}_{l,u};z^{t}_{l,u},(\tilde{\sigma}^{t,i}_{w,u})^2)$. The variable $(\tilde{\sigma}^{t,i}_{w,u})^2$ denotes the effective noise including the background noise and interference signals, which is updated in each iteration and the explicit expression is given in (\ref{equ:EM_sigma_w}).
In this section, we consider the ideal scenario where the received signal $\mathbf{y}^{t}_{u}$ can be perfectly aggregated at the CU. The more practical scenario with finite fronthaul capacity shall be discussed in the next section. Thus, we have
\begin{align}
    \widehat{z}^{t,i}_{l,u} &= \frac{\nu^{p,t,i}_{l,u} y^{t}_{l,u} + (\tilde{\sigma}^{t,i}_{w,u})^2 \widehat{p}^{t,i}_{l,u}}{\nu^{p,t,i}_{l,u} + (\tilde{\sigma}^{t,i}_{w,u})^2} \label{equ:Ez}, \\
    \nu^{z,t,i}_{l,u} &= \frac{\nu^{p,t,i}_{l,u} (\tilde{\sigma}^{t,i}_{w,u})^2}{\nu^{p,t,i}_{l,u}+(\tilde{\sigma}^{t,i}_{w,u})^2} \label{equ:Vz}.
\end{align}

Similarly, the posterior probability of effective channel coefficient $x^{t}_{n,u}$ for $n \in \mathcal{N}_{u}$ can be expressed as
\begin{align}\label{equ:Pp_x}
    p(x^{t}_{n,u,}|\{\mathbf{Y}^{\tau}\}_{\tau \in \mathcal{T}_w}) &= \frac{p(x^{t}_{n,u})p(x^{t}_{n,u}|\widehat{r}^{t}_{n,u})}{\int p(x^{t}_{n,u})p(x^{t}_{n,u}|\widehat{r}^{t}_{n,u}) d x^{t}_{n,u}}, 
\end{align}
where $p(x^{t}_{n,u})$ is assumed to satisfy the independent Bernoulli-Gaussian distribution and we have the conditional probability $p(x^{t}_{n,u}|\widehat{r}^{t,i}_{n,u}) = \mathcal{CN}(x^{t}_{n,u};\widehat{r}^{t,i}_{n,u},\nu^{r,t,i}_{n,u})$. Then the MMSE estimator and the corresponding variance in Lines 16-17 are given by
\begin{align}
    \widehat{x}^{t,i+1}_{n,u} &= \varpi^{t,i}_{n,u} \gamma^{t,i}_{n,u}, \label{equ:Ex} \\
    \nu^{x,t,i+1}_{n,u} &= \varpi^{t,i}_{n,u}[(1-\varpi^{t,i}_{n,u})|\gamma^{t,i}_{n,u}|^2+\nu^{\gamma,t,i}_{n,u}], \label{equ:Dx}
\end{align}
where
\begin{align}
    \varpi^{t,i}_{n,u} &= \left( 1+\frac{1-\overrightarrow{\phi}^{t,i}_{n,u}}{\overrightarrow{\phi}^{t,i}_{n,u}} \frac{g_{n,u}+\nu^{r,t,i}_{n,u}}{\nu^{r,t,i}_{n,u}} \exp  (-\Xi^{t,i}_{n,u} ) \right)^{-1}, \\
    \gamma^{t,i}_{n,u} &= g_{n,u}\cdot(g_{n,u}+\nu^{r,t,i}_{n,u})^{-1} \cdot\widehat{r}^{t,i}_{n,u}, \\
    \nu^{\gamma,t,i}_{n,u} &= \nu^{r,t,i}_{n,u} \cdot g_{n,u} \cdot (g_{n,u}+\nu^{r,t,i}_{n,u})^{-1}, \\
    \Xi^{t,i}_{n,u} &= \left[(\nu^{r,t,i}_{n,u})^{-1} - (\nu^{r,t,i}_{n,u} + g_{n,u})^{-1}\right] \cdot |\widehat{r}^{t,i}_{n,u}|^2.
\end{align}

After the GAMP-based CE, the extrinsic message passed for activity refinement in the next iteration is given by
\begin{align}\label{equ:M_x_to_k}
    \xi^{i+1}_{x^{t}_{n,u} \rightarrow k^{t}_{n}}(\lambda^{t}_{n}) &\propto \prod_{l=1}^{L} \xi^{i}_{f^{t}_{l,u} \rightarrow x^{t}_{n,u}}(x^{t}_{n,u}) \notag \\
    \quad &= (1-\lambda^{t}_{n})(1-\overleftarrow{\phi}^{t,i+1}_{n,u}) + \lambda^{t}_{n} \overleftarrow{\phi}^{t,i+1}_{n,u},
\end{align}
where
\begin{align}\label{equ:leftarrow_p}
    \overleftarrow{\phi}^{t,i+1}_{n,u} &= \frac{p(\widehat{r}^{t}_{n,u}|\lambda^{t}_{n} = 1)}{p(\widehat{r}^{t}_{n,u}|\lambda^{t}_{n} = 0) + p(\widehat{r}^{t}_{n,u}|\lambda^{t}_{n} = 1)} \notag \\
    \quad &= \frac{\mathcal{CN}(\widehat{r}^{t}_{n,u};0,\nu^{r,t,i}_{n,u}+g_{n,u})}{\mathcal{CN}(\widehat{r}^{t}_{n,u};0,\nu^{r,t,i}_{n,u})+ \mathcal{CN}(\widehat{r}^{t}_{n,u};0,\nu^{r,t,i}_{n,u}+g_{n,u})} \notag \\
    \quad &= \left(1+\frac{\nu^{r,t,i}_{n,u}+g_{n,u}}{\nu^{r,t,i}_{n,u}}\exp(-\Xi^{t,i}_{n,u})\right)^{-1}.
\end{align}

\subsubsection{Effective Noise Variance Learning}

Under the user-centric AP cooperation strategy, the signals from user $n \notin \mathcal{N}_{u}$ can be treated as interference at AP $u$. Thus, we propose to learn the statistical parameters of the effective noise including the interference from all users $n \notin \mathcal{N}_{u}$ and the background noise.
Specifically, the received signals at each AP $u$ in the $t$th frame can be rewritten as
\begin{align}\label{equ:rew_yu}
    \mathbf{y}^{t}_{u} &=
     \mathbf{A}_{\mathcal{N}_{u}} \mathbf{x}^{t}_{\mathcal{N}_{u}} + \mathbf{A}_{\bar{\mathcal{N}}_{u}} \mathbf{x}^{t}_{\bar{\mathcal{N}}_{u}} + \mathbf{w}^{t}_{u} = \breve{\mathbf{z}}^{t}_{u} + \bar{\mathbf{z}}^{t}_{u} + \mathbf{w}^{t}_{u} = \breve{\mathbf{z}}^{t}_{u} + \breve{\mathbf{w}}^{t}_{u},
\end{align}
where $\bar{\mathcal{N}}_{u} = \mathcal{N}_0 \setminus \mathcal{N}_{u}$ denote the user set generated by removing $\mathcal{N}_u$ from the entire user set $\mathcal{N}_0$; $\bar{\mathbf{z}}^{t}_{u} = \mathbf{A}_{\bar{\mathcal{N}}_{u}} \mathbf{x}^{t}_{\bar{\mathcal{N}}_{u}}$ is the interference generated from the users in $\bar{\mathcal{N}}_{u}$; $\breve{\mathbf{w}}^{t}_{u} = \bar{\mathbf{z}}^{t}_{u} + \mathbf{w}^{t}_{u}$ is the effective noise vector.
For simplicity, we approximate the interference signal by the Gaussian distribution, which leads to the Gaussian distributed effective noise $\breve{\mathbf{w}}^{t}_{u}$ with zero mean and variance $(\tilde{\sigma}^{t}_{w,u})^2$ at each AP $u$. Here, the effective noise variance is set to $(\tilde{\sigma}^{t,1}_{w,u})^2 = \sigma^2_w$ in the beginning of the algorithm under the assumption that $\sigma^2_w$ is known in advance. Then we update $(\tilde{\sigma}^{t,i+1}_{w,u})^2$ by adopting the expectation maximization (EM) algorithm \cite{Moon_1996_SPM} in each iteration, which adaptively tunes $(\tilde{\sigma}^{t,i+1}_{w,u})^2$ at each AP $v$ from its received signals $\{\mathbf{y}^{t}_{u}\}_{t\in\mathcal{T}_w}$ as
\begin{align}\label{equ:EM_sigma_w}
  (\tilde{\sigma}^{t,i+1}_{w,u})^2 =&~ \frac{1}{LT}\sum_{t\in\mathcal{T}_w} \sum_{l=1}^{L} \Big[|y^{t}_{l,u}-\widehat{z}^{t,i}_{l,u}|^2+\nu^{z,t,i}_{l,u}\Big].
\end{align}

\begin{algorithm}\label{Alg:DCS-MMV-GAMP} \footnotesize
    \caption{The DCS-MMV-GAMP Algorithm for Joint AD and CE}
    \SetKwInOut{Input}{Input}
    \SetKwInOut{Output}{Output}
    \SetKwInOut{Initialize}{Initialize}


    \Input{Received signals $\{\mathbf{Y}^{t}\}_{t\in\mathcal{T}_w}$, pilot matrix $\mathbf{A}$, transmit power $\rho_0$, large-scale attenuations $\{g_{n,u}\}_{n=1,u=1}^{N,U}$, probabilities $\{\alpha,\beta,p_a\}$, noise variance $\sigma^2_w$, detection sets $\{\mathcal{N}_{u}\}_{u=1}^{U}$, connection sets $\{\mathcal{U}_{n}\}_{n\in\mathcal{N}}$, target sub-window $\mathcal{T}_f$, tolerance $\epsilon$, and maximal iteration number $I_{\textrm{max}}$.}
    \BlankLine
    \Output{Estimated effective channels $\{\widehat{\mathbf{X}}^{t}\}_{t\in\mathcal{T}_f}$, estimated probabilities $\{\textbf{Pr}(\lambda^{t}_{n}|\{\mathbf{Y}^{\tau}\}_{\tau\in\mathcal{T}_w})\}_{t = t_1,n=1}^{t_1+\Delta_w-1,N}$. }
    \BlankLine
    \Initialize{$i \leftarrow 1$; $\forall t,n,u: \widehat{x}^{t,i}_{n,u} = 0$, $\nu^{x,t,i}_{n,u} = p_n g_{n,u}$, $\overleftarrow{\phi}^{t,i}_{n,u}=0.5$; $\forall t, l, u: \widehat{s}^{t,i-1}_{l,u} = 0$; $\forall t,u, (\tilde{\sigma}^{t,i}_{w,u})^2 = \sigma^2_w$;}
    \BlankLine

    \While{$\frac{\sum_{t\in\mathcal{T}_w}||\mathbf{X}^{t,i}-\mathbf{X}^{t,i-1}||^2_F}{\sum_{t\in\mathcal{T}_w}||\mathbf{X}^{t,i-1}||^2_F} \ge \epsilon$ {\rm \bf and} $i \le I_{\text{\rm max}}$}{

    \BlankLine
    \{Activity refinement via MP\}

    \nl $\forall t, n: \overleftarrow{\pi}^{t,i}_{n}$ is updated by (\ref{equ:leftarrow_pi});
    \BlankLine

    $t \leftarrow t_0$;

    \While{$t \le (t_0+T_w-1)$}{
    \nl $\forall n: \overrightarrow{\psi}^{t,i}_{n}$ is updated by (\ref{equ:rightarrow_psi});

    \nl $\forall n: \overleftarrow{\varphi}^{t,i}_{n}$ is updated by (\ref{equ:leftarrow_varphi});

    $t \leftarrow t+1$
    }

    $t \leftarrow (t_0+T_w-1)$

    \While{$t \ge t_0$}{
    \nl $\forall n: \overleftarrow{\psi}^{t,i}_{n}$ is updated by (\ref{equ:leftarrow_psi});

    \nl $\forall n: \overrightarrow{\varphi}^{t,i}_{n}$ is updated by (\ref{equ:rightarrow_varphi});

    $t \leftarrow t-1$;
    }

    \nl $\forall t,n,u \in \mathcal{U}_{n}: \overrightarrow{\pi}^{t,i}_{n,u}$ is updated by (\ref{equ:P_lam_to_k});

    \nl$\forall t,n, u \in \mathcal{U}_{n}: \overrightarrow{\phi}^{t,i}_{n,u}$ is updated by (\ref{equ:P_k_to_x});

    \BlankLine
    \{CE via GAMP\}

    \nl $\forall t,l,u: \nu^{p,t,i}_{l,u} = \sum_{n \in \mathcal{N}_{u}}|a_{l,n}|^2 \nu^{x,t,i}_{n,u}$;

    \nl $\forall t,l,u: \widehat{p}^{t,i}_{l,u} = \sum_{n \in \mathcal{N}_{u}} a_{l,n}\widehat{x}^{t,i}_{n,u} - \nu^{p,t,i}_{l,u}\widehat{s}^{t,i-1}_{l,u}$;

    \nl $\forall t,l,u: \nu^{z,t,i}_{l,u} = \mathbb{V}[z^{t}_{l,u}|\widehat{p}^{t,i}_{l,u},\nu^{p,t,i}_{l,u},y^{t}_{l,u},(\tilde{\sigma}^{t,i}_{w,u})^2]$;

    \nl $\forall t,l,u: \widehat{z}^{t,i}_{l,u} = \mathbb{E}[z^{t}_{l,u}|\widehat{p}^{t,i}_{l,u},\nu^{p,t,i}_{l,u},y^{t}_{l,u},(\tilde{\sigma}^{t,i}_{w,u})^2]$;

    \nl $\forall t,l,u: \nu^{s,t,i}_{l,u} = (1-\nu^{z,t,i}_{l,u}/\nu^{p,t,i}_{l,u})/\nu^{p,t,i}_{l,u}$;

    \nl $\forall t,l,u: \widehat{s}^{t,i}_{l,u} = (\widehat{z}^{t,i}_{l,u} - \widehat{p}^{t,i}_{l,u})/\nu^{p,t,i}_{l,u}$;
    \vspace{0.1cm}

    \nl $\forall t,u,n \in\mathcal{N}_u: \nu^{r,t,i}_{n,u} = 1/(\sum_{l=1}^{L}|a_{l,n}|^2 \nu^{s,t,i}_{l,u})$;

    \nl $\forall t,u,n \in \mathcal{N}_u: \widehat{r}^{t,i}_{n,u} = \widehat{x}^{t,i}_{n,u} + \nu^{r,t,i}_{n,u}\sum_{l=1}^{L} a^{*}_{l,n} \widehat{s}^{t,i}_{l,u}$;

    \nl $\forall t,u,n \in \mathcal{N}_u: \nu^{x,t,i+1}_{n,u} = \mathbb{V}[x^{t,i}_{n,u}|\widehat{r}^{t,i}_{n,u},\nu^{r,t,i}_{n,u},\overrightarrow{\phi}^{t,i}_{n,u}]$;

    \nl $\forall t,u,n \in \mathcal{N}_u: \widehat{x}^{t,i+1}_{n,u} = \mathbb{E}[x^{t,i}_{n,u}|\widehat{r}^{t,i}_{n,u},\nu^{r,t,i}_{n,u},\overrightarrow{\phi}^{t,i}_{n,u}]$;

    \nl $\forall t,u,n \in \mathcal{N}_u: \overleftarrow{\phi}^{t,i+1}_{n,u}$ is updated by (\ref{equ:leftarrow_p});

    \BlankLine
    \{Effective noise variance learning via EM\}

    \nl $\forall u: (\tilde{\sigma}^{t,i+1}_{w,u})^2$ is updated by (\ref{equ:EM_sigma_w});

    \BlankLine
    $i \leftarrow i+1$;
    }

    \nl $\forall n, t \in \mathcal{T}_f: \textbf{Pr}(\lambda^{t}_{n} | \{\mathbf{Y}^{\tau}\}_{\tau\in\mathcal{T}_w})$ is updated by (\ref{equ:Pp_lam}).
\end{algorithm}

\subsection{Activity Decision Rule}

After the DCS-MMV-GAMP algorithm converges, the posterior probability of $\lambda^{t}_{n}$ for $t \in \mathcal{T}_f$ based on the standard MP principle is obtained by
\begin{align}\label{equ:Pp_lam}
    &\mathbf{Pr}(\lambda^{t}_{n} |\{\mathbf{Y}^{\tau}\}_{\tau\in\mathcal{T}_w})  \notag \\
    \propto&~ \xi^{\infty}_{k^{t}_{n} \rightarrow \lambda^{t}_{n}}(\lambda^{t}_{n}) \cdot \xi^{\infty}_{q^{t}_{n} \rightarrow \lambda^{t}_{n}}(\lambda^{t}_{n}) \cdot \xi^{\infty}_{q^{t+1}_{n} \rightarrow \lambda^{t}_{n}}(\lambda^{t}_{n}) \notag \\
    \propto&~ (1-\lambda^{t}_{n})\left[(1-\overleftarrow{\pi}^{t,\infty}_{n})(1-\overrightarrow{\psi}^{t,\infty}_{n})(1-\overrightarrow{\varphi}^{t,\infty}_{n})\right]  \notag \\
    &+ \lambda^{t}_{n} \left[\overleftarrow{\pi}^{t,\infty}_{n} \overrightarrow{\psi}^{t,\infty}_{n} \overrightarrow{\varphi}^{t,\infty}_{n}\right],
\end{align}
where $\xi^{\infty}_{k^{t}_{n} \rightarrow \lambda^{t}_{n}}(\lambda^{t}_{n})$, $\xi^{\infty}_{q^{t}_{n} \rightarrow \lambda^{t}_{n}}(\lambda^{t}_{n})$ and $\xi^{\infty}_{q^{t+1}_{n} \rightarrow \lambda^{t}_{n}}(\lambda^{t}_{n})$ are the converged messages; $\overleftarrow{\pi}^{t,\infty}_{n}$, $\overrightarrow{\psi}^{t,\infty}_{n}$ and $\overrightarrow{\varphi}^{t,\infty}_{n}$ are the updated probabilities of activities with respect to these three converged messages.
The activities of all users in the target frame are usually decided based on log likelihood ratio (LLR) test. Specifically, given the hypothesis testing problem
\begin{align}\label{equ:HTP}
    \left\{\begin{array}{l}
            H_0: \lambda^{t}_{n} = 0, \text{user $n$ is inactive in the $t$th frame}, \\
            H_1: \lambda^{t}_{n} = 1, \text{user $n$ is active in the $t$th frame},
          \end{array} \right.
\end{align}
the Bayes-Risk decision rule can be expressed as
\begin{small}
\begin{align}\label{equ:AD}
    l^t_{n} =&~ \log\left( \frac{\text{Pr}\Big(\lambda^{t}_{n} = 1|\{\mathbf{Y}^{\tau}\}_{\tau \in \mathcal{T}_w}\Big)}{\text{Pr}\Big(\lambda^{t}_{n} = 0|\{\mathbf{Y}^{\tau}\}_{\tau \in \mathcal{T}_w}\Big)} \right) \notag \\
    \quad =&~ \log\left( \frac{\overleftarrow{\pi}^{t,\infty}_{n} \overrightarrow{\psi}^{t,\infty}_{n} \overrightarrow{\varphi}^{t,\infty}_{n}}{\Big(1-\overleftarrow{\pi}^{t,\infty}_{n}\Big) \Big(1-\overrightarrow{\psi}^{t,\infty}_{n}\Big) \Big(1-\overrightarrow{\varphi}^{t,\infty}_{n}\Big)} \right) \mathop{\gtrless}\limits_{H_0}^{H_1} \iota, 
\end{align}
\end{small}where $\iota$ is the decision threshold that can be determined by empirical results. In contrast to the traditional activity detector in \cite{Chen_2019_TWC, Ke_2021_JSAC}, it is observed that all the signals in the window $\mathcal{T}_w$ have an essential influence on the activity decision in the target frame.

\begin{remark}
If we have no knowledge of the temporally correlated activity and simply set $p_a = \alpha = \beta$, we will always have $\overrightarrow{\pi}_{n}^{t,i} = p_a, \forall t,n,i$, and thus no useful information is conveyed between the adjacent frames. Our proposed DCS-MMV-GAMP algorithm will only account the spatial correlation of the effective channels and then reduce to the MP-based algorithm in \cite{Zou_2020_SPL}, denoted as CS-MMV-GAMP.
As such, the proposed DCS-MMV-GAMP algorithm can be considered as a general algorithm framework for joint AD and CE in massive access, which can exploit the spatial-temporal correlation in the signals.
\end{remark}

\begin{remark}
The proposed DCS-MMV-GAMP algorithm can also be employed in the system where each AP is equipped with multiple antennas. By regarding each antenna as one virtual single-antenna AP, the proposed algorithm can be directly utilized without any modifications.
\end{remark}


\subsection{Convergence Analysis}

We utilize the state evolution (SE) to analyze the convergence of the proposed DCS-MMV-GAMP algorithm under the large-scale system limit. When the elements in the pilot matrix satisfy i.i.d. sub-Gaussian distributions, the MSE performance of AMP-based algorithms can be accurately tracked by a set of SE equations in the asymptotic regime where $L, N \rightarrow \infty$ but $L/N$ is fixed \cite{Bayati_2011_TIT,Rangan_2011_ISIT}.

Following \cite{Bayati_2011_TIT,Rangan_2011_ISIT}, we consider that DCS-MMV-GAMP has scalar variances for each frame $t$ and each AP $u$. Similar to our previous work \cite{Zhu_2023_TCOM}, these scalar variances at each iteration $i$ can be defined as
\begin{align}
  \nu^{p,t,i}_{u} &= \frac{|\mathcal{N}_u|}{L} \nu^{x,t,i}_{u}, ~\forall t,u,i, \label{equ:sv_p} \\
  \nu^{z,t,i}_{u} &= \frac{1}{L} \sum_{l=1}^{L} \mathbb{V}[z^{l,t}|\widehat{p}^{t,i}_{l,u},\nu^{p,t,i}_{u},y^t_{l,u},(\tilde{\sigma}^{t,i}_{w,u})^2], ~\forall t,u,i, \label{equ:sv_z} \\
  \nu^{s,t,i}_{u} &= \frac{\nu^{p,t,i}_{u} - \nu^{z,t,i}_{u}}{(\nu^{p,t,i}_{u})^2}, ~\forall t,u,i, \label{equ:sv_s} \\
  \nu^{r,t,i}_{u} &= (\nu^{s,t,i}_{u})^{-1}, ~\forall t,u,i, \label{equ:sv_r}
\end{align}
where we have $\nu^{x,t,i+1}_{u} = \frac{1}{|\mathcal{N}_u|}\sum_{n \in \mathcal{N}_u} \nu^{x,t,i+1}_{n,u}$ with $\nu^{x,t,i+1}_{n,u}$ given in (\ref{equ:Dx}).
Then for the SE, the asymptotic MSE for each frame $t$ and each AP $u$ can be tracked based on the following recursion \cite{Rangan_2011_ISIT,Kamilov_2012_TSP} as
\begin{subequations}\label{equ:se}
\begin{align}
  \bar{\nu}^{x,t,i}_{u} &= \frac{1}{|\mathcal{N}_u|} \sum_{n \in \mathcal{N}_u} \mathbb{E}\left\{\mathbb{V}[x^{t,i}_{n,u}|\widehat{r}^{t,i}_{n,u},\bar{\nu}^{r,t,i}_{u},\overrightarrow{\phi}^{t,i}_{n,u}]\right\}. \label{equ:se_x} \\
  \bar{\nu}^{r,t,i}_{u} &= \mathbb{E}^{-1}\Bigg[\frac{\nu^{p,t,i}_{u} - \nu^{z,t,i}_{u}}{(\nu^{p,t,i}_{u})^2}\Bigg], \label{equ:se_r}
\end{align}
\end{subequations}
where $\nu^{z,t,i}_{u}$ is obtained by (\ref{equ:sv_z}).
The above recursion is initialized with $\bar{\nu}^{x,t,i}_{u} = \frac{\sum_{n\in\mathcal{N}_u}p_n\rho_0g_{n,u}}{|\mathcal{N}_u|}$.
Here, the expectation in (\ref{equ:se_x}) is calculated over $\{ \widehat{r}^{t,i}_{n,u} \}_{n \in \mathcal{N}_u}$ and the expectation in (\ref{equ:se_r}) is calculated over $\{\widehat{p}^{t,i}_{l,u}\}_{l=1}^{L}$ and $\breve{\mathbf{w}}^{t}_{u}$.
However, the probability $\overrightarrow{\phi}^{t,i}_{n,u}$ needs to be updated in each iteration and the explicit expression of the marginal probability on $\overrightarrow{\phi}^{t,i}_{n,u}$ is unavailable. To tackle this problem, we resort to the Monte Carlo method to obtain $\overrightarrow{\phi}^{t,i}_{n,u}$ in each iteration and then substitute it into (\ref{equ:se_x}). Similarly, the effective noise variance $(\tilde{\sigma}^{t,i}_{w,u})^2$ in each iteration is also obtained by the Monte Carlo method. Finally, the asymptotic MSE of the estimation for $\mathbf{X}^t$ in the $i$th iteration can be obtained as $\text{MSE}(\widehat{\mathbf{X}}^{t,i}) = \frac{1}{U}\sum_{u=1}^{U} \bar{\nu}^{x,t,i}_{u}$.

Based on the SE, the asymptotic MSE in each iteration of the proposed scheme can be theoretically derived, which then allows us to analyze the convergence of the proposed scheme under different system settings.
In Section VI-B, we will numerically validate our theoretical analysis, which demonstrates that our scheme guarantees good convergence performance.

\subsection{Distributed Implementation}


\begin{figure}[t]
  \centering
  \subfigure[The DCS-MMV-GAMP algorithm]
  {\includegraphics[width=.38\textwidth]{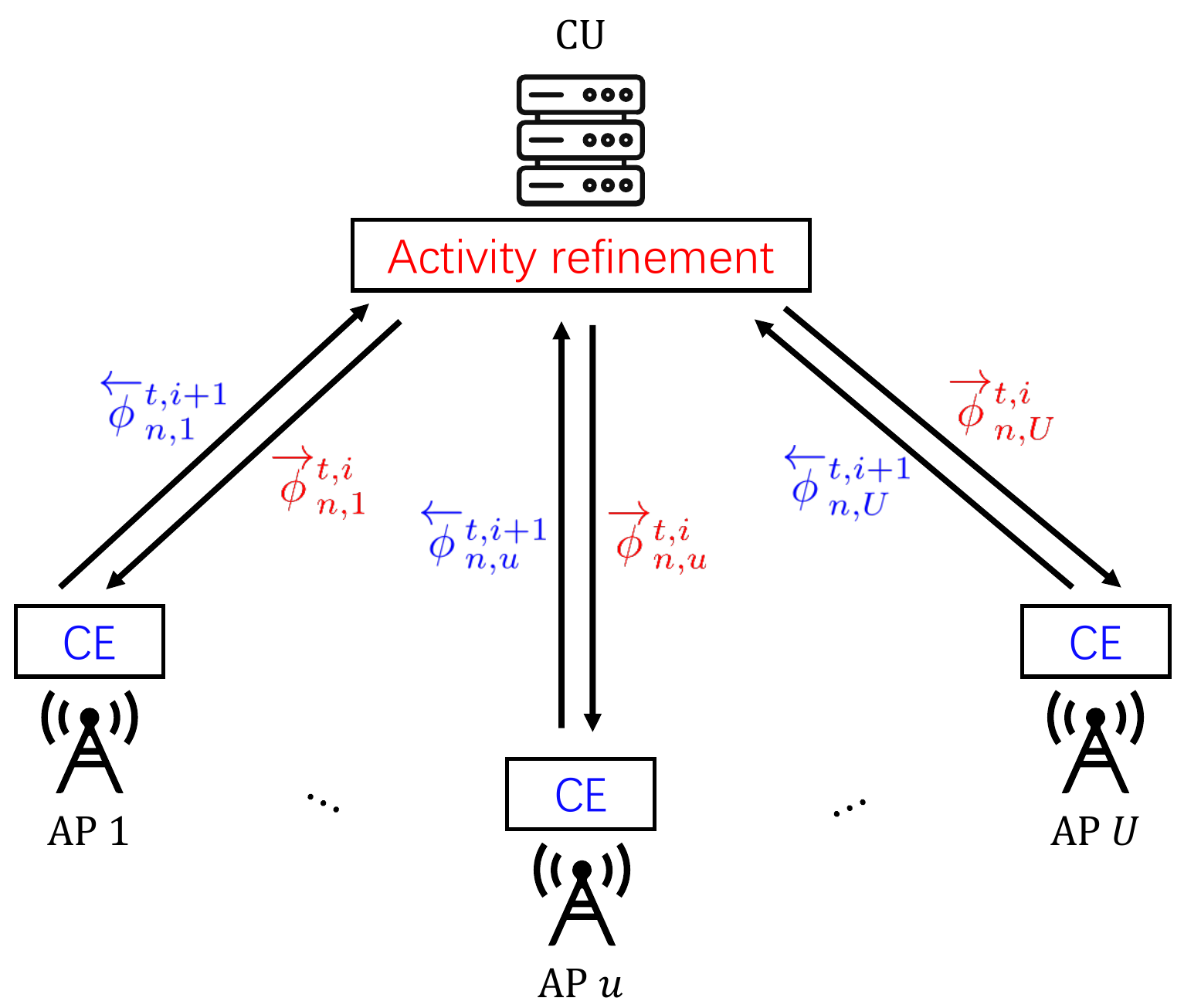}}\hspace{1cm}
  \subfigure[The vAMP-based algorithms]
  {\includegraphics[width=.424\textwidth]{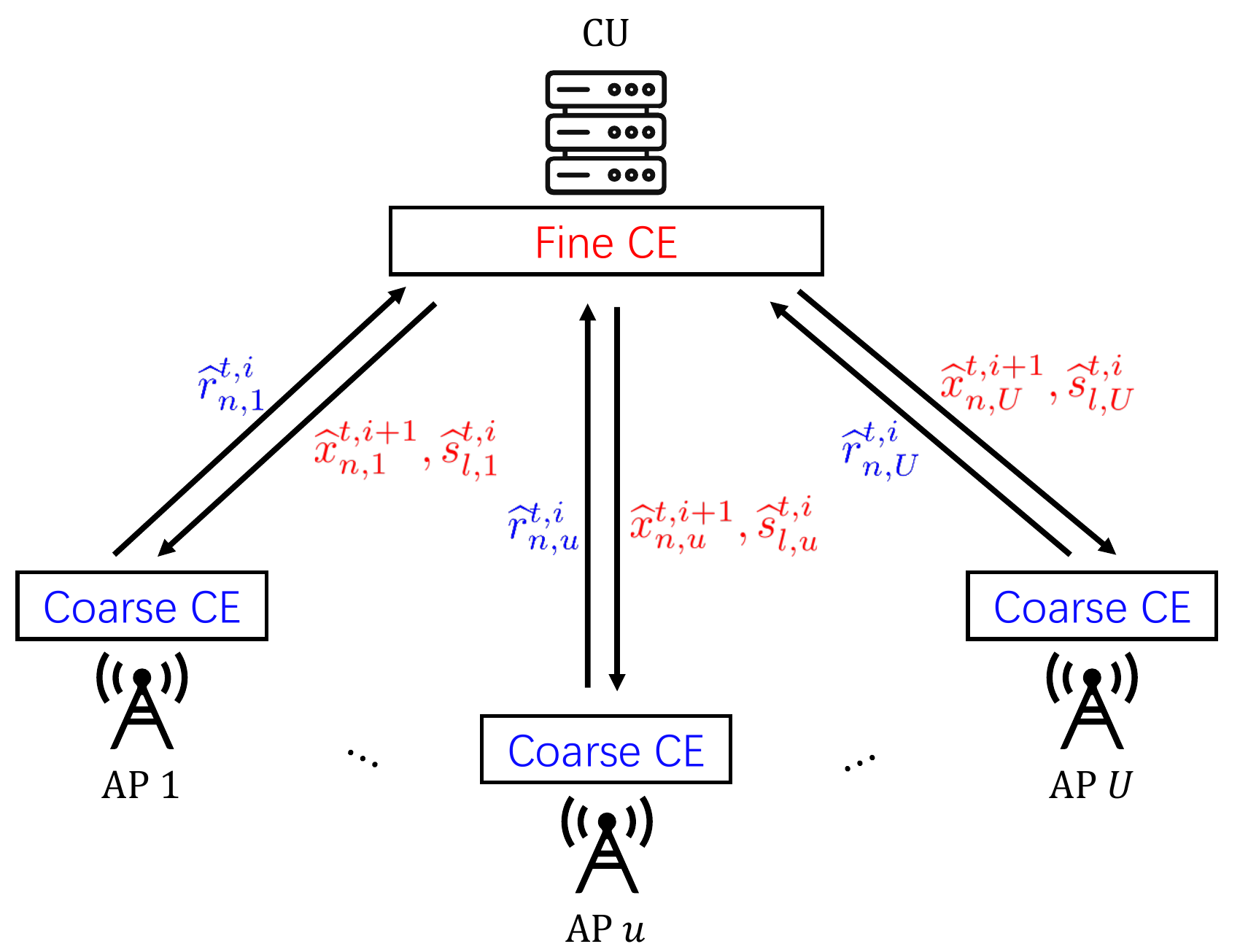}}
  \vspace{-0.3cm}
  \caption{The distributed implementation of the joint AD and CE algorithms with infinite fronthaul capacity.}\label{Fig:D_MP}
  \vspace{-0.6cm}
\end{figure}

In the C-RAN, the CU may have the computational burden when the network size becomes very large. Thus, the distributed implementation of the joint AD and CE algorithms is attractive for large-scale networks, where part of the computations are offloaded to the APs to alleviate the computational cost at the CU.
When the fronthaul capacity is very large, the functional splits in the distributed implementations of the DCS-MMV-GAMP algorithm and the conventional vector shrinkage function-equipped AMP (vAMP)-based algorithms \cite{Chen_2019_TWC, Wang_2021_ISIT} can be quite different, which are shown in Fig. \ref{Fig:D_MP}.
For DCS-MMV-GAMP, the CE part is employed at each AP and the activity refinement part is performed at the CU. In the $i$th iteration, the refined probabilities $\overleftarrow{\phi}^{t,i+1}_{n,u}$ and $\overrightarrow{\phi}^{t,i}_{n,u}$ are exchanged via the fronthaul link.
While for the vAMP-based algorithms, each AP transmits the noise-corrupted channel $\widehat{r}^{t,i}_{n,u}$ via coarse CE to the CU, then the CU transmits the fine-estimated channel $\widehat{x}^{t,i}_{n,u}$ output by the vector shrinkage function as well as the residual signal $\widehat{s}^{t,i}_{l,u}$ back to each AP in each iteration.
Note that since the size of the active probabilities $\overrightarrow{\phi}^{t,i}_{n,u}$ is usually much smaller than that of  $\{\widehat{\mathbf{X}}^t_u\}_{t\in\mathcal{T}_w}$ with $\widehat{\mathbf{X}}^t_u = [\widehat{\mathbf{x}}^t_{u,1},\dots,\widehat{\mathbf{x}}^t_{u,M}] \in \mathbb{C}^{N \times M}$ if $M\ge2$ antennas are equipped at each AP, the proposed DCS-MMV-GAMP algorithm is thus more favorable for distributed implementation.

However, when the fronthaul capacity is limited, the frequent content exchange in each iteration of these two algorithms will be hindered. As a result, these two kind of algorithms can only perform local estimation and then the CU makes the final activity detection decision by aggregating the local estimations from each AP. The details of the distributed implementations in this case will be introduced in the next section.

\subsection{Computational Complexity Analysis}

This subsection provides the computational complexity of the proposed algorithm and compare it with the baseline algorithms, i.e., vAMP \cite{Chen_2018_TSP,Liu_2018_TSP}, SI-aided MMV-AMP \cite{Wang_2021_ISIT}, and CS-MMV-GAMP \cite{Zou_2020_SPL} in Table \ref{table:CC}. We consider the general case where each AP has $M \ge 1$ antennas. The user-centric AP cooperation strategy is employed in the CS-MMV-GAMP algorithm and the proposed DCS-MMV-GAMP algorithm, while the vAMP algorithm and the SI-aided MMV-AMP algorithm consider that each AP will detect all the active users in the network, since their specific implementations with the user-centric AP cooperation strategy has not been designed yet. To fairly compare their performance, we simply set $N_d = N$ and $U_d = U$. The main computational cost of all the algorithms comes from AMP or GAMP, which scales linearly with $L, N, M, U$. Since DCS-MMV-GAMP needs to estimate the channels of all the frames in the sliding window, its computational complexity also scales linearly with the factor $\frac{T_w}{|\mathcal{T}_f|}$. Based on the first-order Markov process, the factor $\frac{T_w}{|\mathcal{T}_f|}$ usually stays in the range of $[1,5]$ by our simulation trials. Thus we can carefully design $\mathcal{T}_w$ and $\mathcal{T}_f$ to reduce the average computational cost without degrading the performance. As will be shown in Section \ref{sec:simulation}, the additional computational cost enables the proposed algorithm to better exploit the temporally correlated activity. Assuming that the distance between two adjacent APs is $2r_0$, we can obtain $N_d = \frac{\pi D_{max}^2 N}{2\sqrt{3}r_0^2U}$ and $V_d = \frac{\pi D_{max}^2}{2\sqrt{3}r_0^2}$ under the user-centric AP cooperation strategy. Therefore, in the large-scale network where $N$ and $U$ is very large but with fixed $\frac{N}{U}$, we have $N_d \ll N$ and $U_d \ll U$, meaning that the computational complexity can be greatly reduced.

\begin{table*}[t]
  \centering
  \caption{Computational Complexity Comparison}\label{table:CC}
    \begin{tabular}{c|c}
    \hline
    \hline
    \textbf{Algorithm} & Average number of complex multiplications in each iteration \\
    \hline
    vAMP & $2LNUM + LU^2M^2 + NU^2M^2 + \frac{7}{4}NUM + LUM + \frac{25}{4}N$ \\
    SI-aided MMV-AMP & $2LNUM + LU^2M^2 + NU^2M^2 + \frac{7}{4}NUM + LUM + \frac{29}{4}N$ \\
    CS-MMV-GAMP & $2LN_d UM + \frac{17}{4}LUM + \frac{27}{4}N_d UM + (\frac{1}{2}U_d+1)N$ \\
    DCS-MMV-GAMP & $\frac{T_w}{|T_f|}\left[2LN_d UM + \frac{17}{4}LUM + \frac{27}{4}N_d UM + (U_d+\frac{67}{4})N\right]$ \\
    \hline
    \hline
	\end{tabular} \\
\vspace{0.2cm}
Note: $N_d = \mathbb{E}[|\mathcal{N}_{u}|]$ indicates the average number of the detected users for each AP $u$, $U_d = \mathbb{E}[|\mathcal{U}_{n}|]$ indicates the average number of APs for the detection of each user $n$.
\vspace{-0.6cm}
\end{table*}

\section{Algorithm Implementation with Finite Fronthaul Capacity}\label{sec:finite_capacity}

In this section, we consider the practical scenario where the fronthaul link has finite capacity. As such, the frequent exchange of activity information between the CU and all APs are unavailable.
To cope with the fronthaul capacity limit, we develop two schemes of QF and DF based on the proposed DCS-MMV-GAMP algorithm depending on the function split of the C-RAN.
In QF, the received signals at each AP $u$ are first quantized with the resolution of $b_Q$ bits per sample and then transmitted to the CU for centralized processing. In DF, each AP $u$ first performs local AD based on its own received signals and then forwards the detected activity information in terms of LLRs with the resolution of $b_D$ bits per sample to the CU for final decision.

\subsection{Quantize-and-Forward}

Under QF, the collected quantized signal at the CU from each AP $u$ in the $t$th frame is given by $\tilde{\mathbf{y}}^{t}_{u} = \mathcal{Q}_c(\mathbf{y}^{t}_{u}) = \mathcal{Q}_c(\mathbf{z}^{t}_{u}+\mathbf{w}^{t}_{u})$.
The complex quantization function $\mathcal{Q}_c(\cdot)$ operates on the real and imaginary parts of each element in $\mathbf{y}^{t}_{v}$ individually, i.e., $\mathcal{Q}_c(y^{t}_{l,u}) = \mathcal{Q}_r\left(\Re\{y^{t}_{l,u}\}\right) + {\rm j} \mathcal{Q}_r\left(\Im \{y^{t}_{l,u}\}\right)$, where $\mathcal{Q}_r(\cdot)$ is the quantization function in the real number field with the resolution of $b^r_y=\frac{b_Q}{2}$ bits per sample. For simplicity, the uniform quantization function is adopted. Let the set of quantized values be denoted as $\{\zeta_{1},\zeta_{2},\dots,\zeta_{b}\}$ and the set of thresholds be denoted as $\{\varrho_{1},\varrho_{2},\dots,\varrho_{b+1}\}$, where we have $b = 2^{b^r_y}$, $\zeta_{1} < \zeta_{2} < \dots < \zeta_{b}$, and $\varrho_{1} < \varrho_{2} < \dots < \varrho_{b+1}$ with $\varrho_{1} = -\infty$ and $\varrho_{b+1} = \infty$. The quantization function will output $\tilde{y}^t_{l,u} = \zeta_{c}+{\rm j}\zeta_{e}$ if the input signal $y^{t}_{l,u}$ falls in the region $\{y^{t}_{l,u}|\varrho_{c} < \Re\{y^{t}_{l,u}\} \le \varrho_{c+1}, \varrho_{e} < \Im\{y^{t}_{l,u}\} \le \varrho_{e+1}\}$. The conditional probability of the output channel can be expressed as
\begin{align}\label{equ:p_yz}
    &~ p(\tilde{y}^{t}_{l,u} = \zeta_{c}+{\rm j}\zeta_{e}|z^{t}_{l,u}) \notag \\
    =&~ \frac{1}{\pi (\tilde{\sigma}^{t,i}_{w,u})^2} \int_{\varrho_{c}}^{\varrho_{c+1}} \int_{\varrho_{e}}^{\varrho_{e+1}} \mathcal{N}(y^R;\Re\{z^{t}_{l,u}\},\frac{(\tilde{\sigma}^{t,i}_{w,u})^2}{2}) \notag \\
    &\times \mathcal{N}(y^I;\Im\{z^{t}_{l,u}\},\frac{(\tilde{\sigma}^{t,i}_{w,u})^2}{2}) dy^R dy^I \notag \\
    =&~ \Bigg[\Phi\Bigg(\frac{\sqrt{2}(\varrho_{c+1} - \Re \{z^{t}_{l,u}\})}{\tilde{\sigma}^{t,i}_{w,u}}\Bigg) - \Phi\Bigg(\frac{\sqrt{2}(\varrho_{c} - \Re \{z^{t}_{l,u}\})}{\tilde{\sigma}^{t,i}_{w,u}}\Bigg)\Bigg] \notag \\
     &~\times \Bigg[\Phi\Bigg(\frac{\sqrt{2}(\varrho_{e+1} - \Im \{z^{t}_{l,u}\})}{\tilde{\sigma}^{t,i}_{w,u}}\Bigg) \notag \\
     &- \Phi\Bigg(\frac{\sqrt{2}(\varrho_{e} - \Im \{z^{t}_{l,u}\})}{\tilde{\sigma}^{t,i}_{w,u}}\Bigg)\Bigg],
\end{align}
where $\Phi(\cdot)$ is the cumulative standard Gaussian distribution function. Since the GAMP method can handle arbitrary distribution of $p(\tilde{y}^{t}_{l,u}|z^{t}_{l,u})$, the proposed DCS-MMV-GAMP algorithm can take the signal quantization into consideration for QF. Specifically, the MMSE estimator of the noiseless signal $z^{t}_{l,u}$ in Line 11 of Algorithm \ref{Alg:DCS-MMV-GAMP} is modified as
\begin{align}\label{equ:Ez_Q}
    \widehat{z}^{t,i}_{l,u} =&~
    \mathbb{E}\left[\Re\{z^{t}_{l,u}\}|\Re\{\widehat{p}^{t,i}_{l,u}\},\frac{\nu^{p,t,i}_{l,u}}{2},\Re\{\tilde{y}^{t}_{l,u}\},\frac{(\tilde{\sigma}^{t,i}_{w,u})^2}{2}\right] \notag \\
    \quad &+ {\rm j} \mathbb{E}\left[\Im\{z^{t}_{l,u}\}|\Im\{\widehat{p}^{t,i}_{l,u}\},\frac{\nu^{p,t,i}_{l,u}}{2},\Im\{\tilde{y}^{t}_{l,u}\},\frac{(\tilde{\sigma}^{t,i}_{w,u})^2}{2}\right] \notag \\
    \quad =&~
    Z\left(\Re\{\widehat{p}^{t,i}_{l,u}\},\frac{\nu^{p,t,i}_{l,u}}{2},\Re\{\tilde{y}^{t}_{l,u}\},\frac{(\tilde{\sigma}^{t,i}_{w,u})^2}{2}\right) \notag \\
    \quad &+ {\rm j} Z\left(\Im\{\widehat{p}^{t,i}_{l,u}\},\frac{\nu^{p,t,i}_{l,u}}{2},\Im\{\tilde{y}^{t}_{l,u}\},\frac{(\tilde{\sigma}^{t,i}_{w,u})^2}{2}\right),
\end{align}
where the function $Z(p,\nu^p,y,\sigma^2)$ is given by
\begin{align}
    Z(p,\nu^p,y,\sigma^2) &= p + \frac{\textrm{sign}(y)\nu^p}{\sqrt{2(\sigma^2+\nu^p)}}\left( \frac{\eta(\kappa_1)-\eta(\kappa_2)}{\Phi(\kappa_1)-\Phi(\kappa_2)} \right), \label{equ:Z} \\
    \kappa_1 &= \frac{\textrm{sign}(y)p-\min(|\varrho_{c}|,|\varrho_{c+1}|)}{\sqrt{\frac{\sigma^2+\nu^p}{2}}}, \label{equ:kappa1} \\
    \kappa_2 &= \frac{\textrm{sign}(y)p-\max(|\varrho_{c}|,|\varrho_{c+1}|)}{\sqrt{\frac{\sigma^2+\nu^p}{2}}}, \label{equ:kappa2}
\end{align}
with $\eta(\kappa) \triangleq \frac{1}{\sqrt{2\pi}}\exp(-\kappa^2/2)$. The derivations of (\ref{equ:Z})-(\ref{equ:kappa2}) are similar to those in \cite{Wen_2016_TSP}.
The variance of the MMSE estimator can also be computed as
\begin{align}\label{equ:Vz_Q}
    \nu^{z,t,i}_{l,u} =&~ \mathbb{V}\left[\Re\{z^{t}_{l,u}\}|\Re\{\widehat{p}^{t,i}_{l,u}\},\frac{\nu^{p,t,i}_{l,u}}{2},\Re\{y^{t}_{l,u}\},\frac{(\tilde{\sigma}^{t,i}_{w,u})^2}{2}\right] \notag \\
    \quad &+  \mathbb{V}\left[\Im\{z^{t}_{l,u}\}|\Im\{\widehat{p}^{t,i}_{l,u}\},\frac{\nu^{p,t,i}_{l,u}}{2},\Im\{y^{t}_{l,u}\},\frac{(\tilde{\sigma}^{t,i}_{w,u})^2}{2}\right] \notag \\
    \quad =&~ V\left(\Re\{\widehat{p}^{t,i}_{l,u}\},\frac{\nu^{p,t,i}_{l,v}}{2},\Re\{y^{t}_{l,u}\},\frac{(\tilde{\sigma}^{t,i}_{w,u})^2}{2}\right) \notag \\
    \quad &+ V\left(\Im\{\widehat{p}^{t,i}_{l,u}\},\frac{\nu^{p,t,i}_{l,u}}{2},\Im\{y^{t}_{l,u}\},\frac{(\tilde{\sigma}^{t,i}_{w,u})^2}{2}\right),
\end{align}
where
\begin{align}
    V(p,\nu^p,y,\sigma^2) =&~ \frac{\nu^{p}}{2} - \frac{(\nu^{p})^2}{2(\sigma^2+\nu^{p})} \Bigg( \frac{\kappa_1\eta(\kappa_1)-\kappa_2\eta(\kappa_2)}{\Phi(\kappa_1)-\Phi(\kappa_2)} \notag \\
    \quad &+\Bigg( \frac{\eta(\kappa_1)-\eta(\kappa_2)}{\Phi(\kappa_1)-\Phi(\kappa_2)} \Bigg)^2 \Bigg).
\end{align}

However, we find that the effective noise variance learning operation can deteriorate the performance of DCS-MMV-GAMP with small $b_Q$ from the simulation trials.
Therefore, in QF, we remove the operation of Line 19 and always let $(\tilde{\sigma}^{t,i+1}_{w,u})^2 = \sigma_w^2, \forall t, u$ at each iteration in Algorithm \ref{Alg:DCS-MMV-GAMP}.

\subsection{Detect-and-Forward}\label{SubSec:DF}

Under DF, the cooperative AD is realized by LLR aggregation at the CU while independent CE is executed in each AP.
In specific, each AP $v$ first employs DCS-MMV-GAMP to obtain the estimated effective channels $\{\widehat{\mathbf{x}}^{t}_{u}\}_{t \in \mathcal{T}_f}$
and the locally detected activity LLRs $\{\pmb{l}^{t}_{u}\}_{t \in \mathcal{T}_f}$ based on the received signals $\{\mathbf{y}^{t}_{u}\}_{t\in\mathcal{T}_w}$, where $\pmb{l}^{t}_{u} = [l^t_{1,u},\dots,l^t_{N,u}]^T$ and $l^r_{n,u}$ has the similar derivation to (\ref{equ:AD}).
The locally detected activity LLRs will be quantized and then forwarded to the CU, where these LLRs are quantized by the same uniform quantization function for convenience, i.e., $l^{Q,t}_{n,u} = \mathcal{Q}_{l}(l^{t}_{n,u}), \forall t,n,u$. At last, the CU adds up the quantized LLRs of each user from its connected APs, i.e., $l^{Q,t}_{n} = \sum_{u \in \mathcal{U}_{n}} l^{Q,t}_{n,u}, \forall t,n$, and then compares the LLR $l^{Q,t}_{n}$ with a predetermined threshold to make the final decision on the activities of all users.

To analyze the performance of these two approaches of QF and DF, we first define the fronthaul capacity as the number of bits $B$ that can be transmitted from each AP to the CU with no error in a given time interval. We then consider the general case where there are $M$ antennas equipped at each AP, such that the received signals at each AP can be represented by $\mathbf{Y}^{t}_{u} = [\mathbf{y}^{t}_{u,1}, \dots, \mathbf{y}^{t}_{u,M}]$. Therefore, we have $B = LMb_Q$ and $B=|\mathcal{N}_{u}|b_D$ for QF and DF, respectively. Though the work \cite{Utkovski_2017_SPL} has considered the two schemes of QF and DF with limited fronthaul capacity, it only concentrates on the single-antenna scenario and thus ignores the fact that the antenna number $M$ also has a great influence on the performance of QF and DF. Thus, besides only considering the impact of $L, N$ on the performance of QF and DF, we also study how the antenna number $M$ affects the performance of these two schemes in this work.
First, when $B$ is large enough, it implies that both $b_Q$ and $b_D$ are also very large. So that QF can nearly achieve the optimal performance of DCS-MMV-GAMP and thus always outperform DF since the AP cooperation is not fully exploited in DF. On the other hand, when $B$ is limited, the performance superiority of QF and DF can be different depending on the antenna number $M$. Specifically if each AP is equipped with a small number of antennas, then we have $b_Q = \frac{b_D|\mathcal{N}_{u}|}{LM} \gg b_D$ since $L \ll N$ is usually assumed in massive access \cite{Chen_2021_JSAC}. As such, QF usually outperforms DF. However, if $M$ is enlarged, $b_Q$ becomes comparable to $b_D$, so that the performance of DF can approach that of QF. By further increasing the antenna number $M$, the quantization resolution $b_Q$ of QF will be further reduced and thus deteriorate the detection performance. However, the quantization resolution $b_D$ in DF is independent of $M$ and more measurements are obtained at each local AP, which can boost the detection performance of DF. As a result, DF may have superior performance to QF.

\begin{remark}
In QF, each element of the received signals is quantized with a scalar uniform quantization function. However, this quantization scheme is not optimal in CS, and thus more advanced quantization schemes \cite{Kamilov_2012_TSP} can be employed to enhance the performance of QF.
In DF, we also consider the simple uniform quantization method. Due to the sporadic traffic, the detected LLRs of most users are much lower than zero. This means that the detected active probabilities of most users approach zero and then the detected active probability vector is in fact a sparse vector. Therefore, each AP $u$ can use advanced lossy compression techniques, such as sparse coding \cite{Wei_2012_TIT} and deep learning \cite{Qin_2019_WCM}, to achieve higher compression efficiency.
\end{remark}

\section{Numerical Results}\label{sec:simulation}

Numerical results are provided in this section to verify the superior performance of the proposed methods. We consider a three-tier network consisting of $U=19$ hexagonal cells as in Fig. \ref{Fig:C-RAN_Model}, where $N_c=1000$ users are uniformly distributed at random in each cell and thus there are total $N=N_cU = 19000$ users in the network. The distance between the adjacent APs is $2r_0 = \sqrt{3}$ km. The path loss between the AP $u$ and the user $n$ at a distance $d_{n,u}$ in kilometer is modeled as $-128.1 - 37.6\log_{10}(d_{n,u})\text{~dB}$ \cite{Chen_2018_TSP}. The transmit power is $\rho_0 = 13$ dBm per user and the background noise power is $-174$dBm/Hz over $10$MHz.
We also consider the system with $\beta = 0.9$ and $p_a = 0.1$, unless otherwise specified. All the numerical results in this section are obtained by averaging $10^4$ numerical trials. We utilize the error detection ratio (EDR) and normalized mean square error (NMSE) as the performance metrics of the AD and CE, respectively. These two performance metrics for the signals at each frame $t$ are defined as
\begin{align}
    \text{EDR}(t)     &= \frac{N^t_e}{N}, \\
    \text{NMSE}(t) &= 10\log_{10}\left(\frac{\sum_{u=1}^{U}\sum_{n\in\mathcal{N}_u}|\widehat{x}^{t}_{n,u} - x^{t}_{n,u}|^2}{\sum_{u=1}^{U}\sum_{n\in\mathcal{N}_u}|x^{t}_{n,u}|^2}\right),
\end{align}
where $N^t_e$ denotes the number of users whose activities at the $t$th frame are wrongly detected.

\begin{figure}[t]
  \centering
  \begin{minipage}[t]{.5\textwidth}
    \center
    \includegraphics[width=0.9\textwidth]{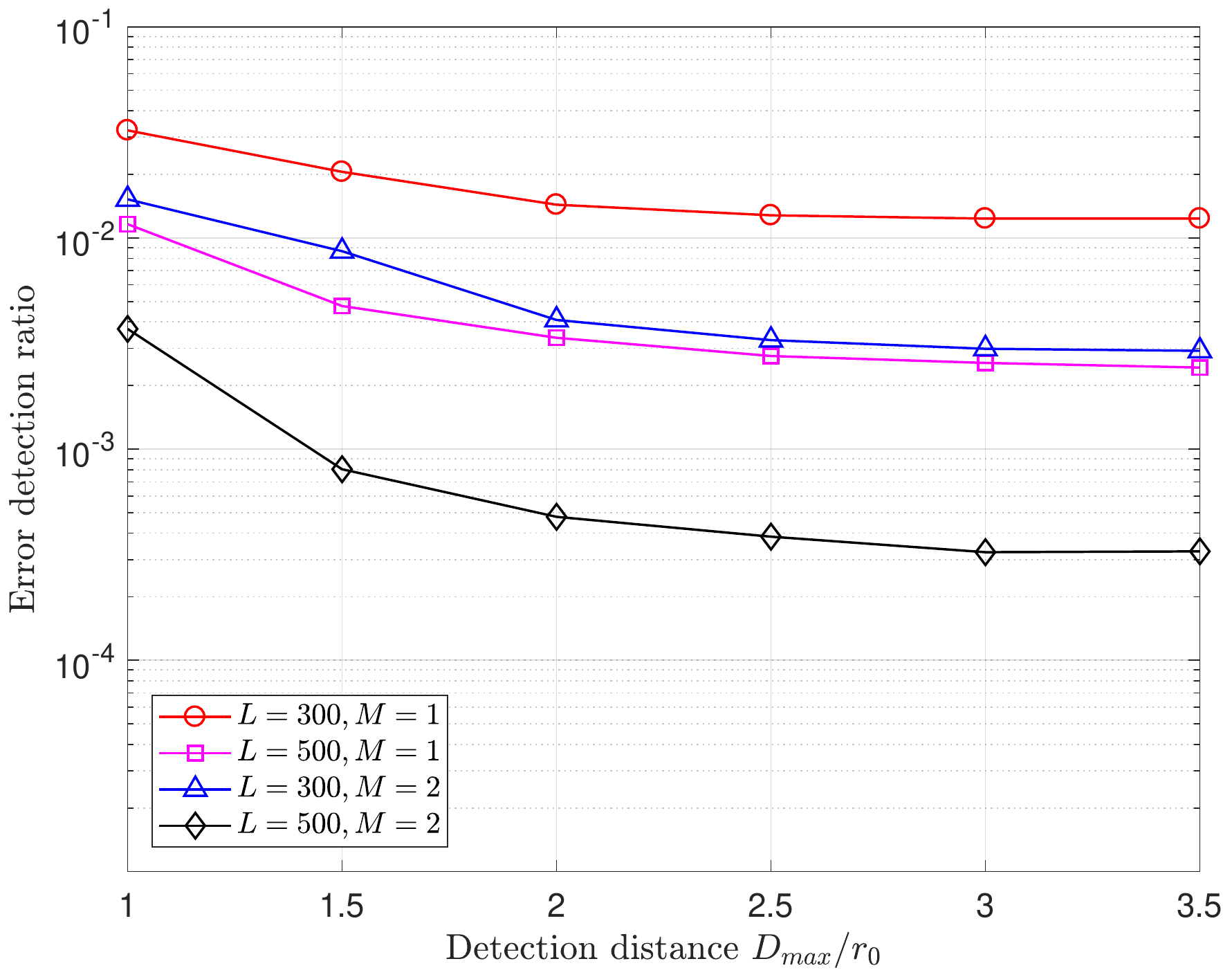}
    \vspace{-0.2cm}
    \caption{The AD performance of DCS-MMV-GAMP under different detection distances.}\label{Fig:EDP_Dth}
    \vspace{0.3cm}
  \end{minipage}
  \begin{minipage}[t]{.5\textwidth}
    \center
    \includegraphics[width=0.9\textwidth]{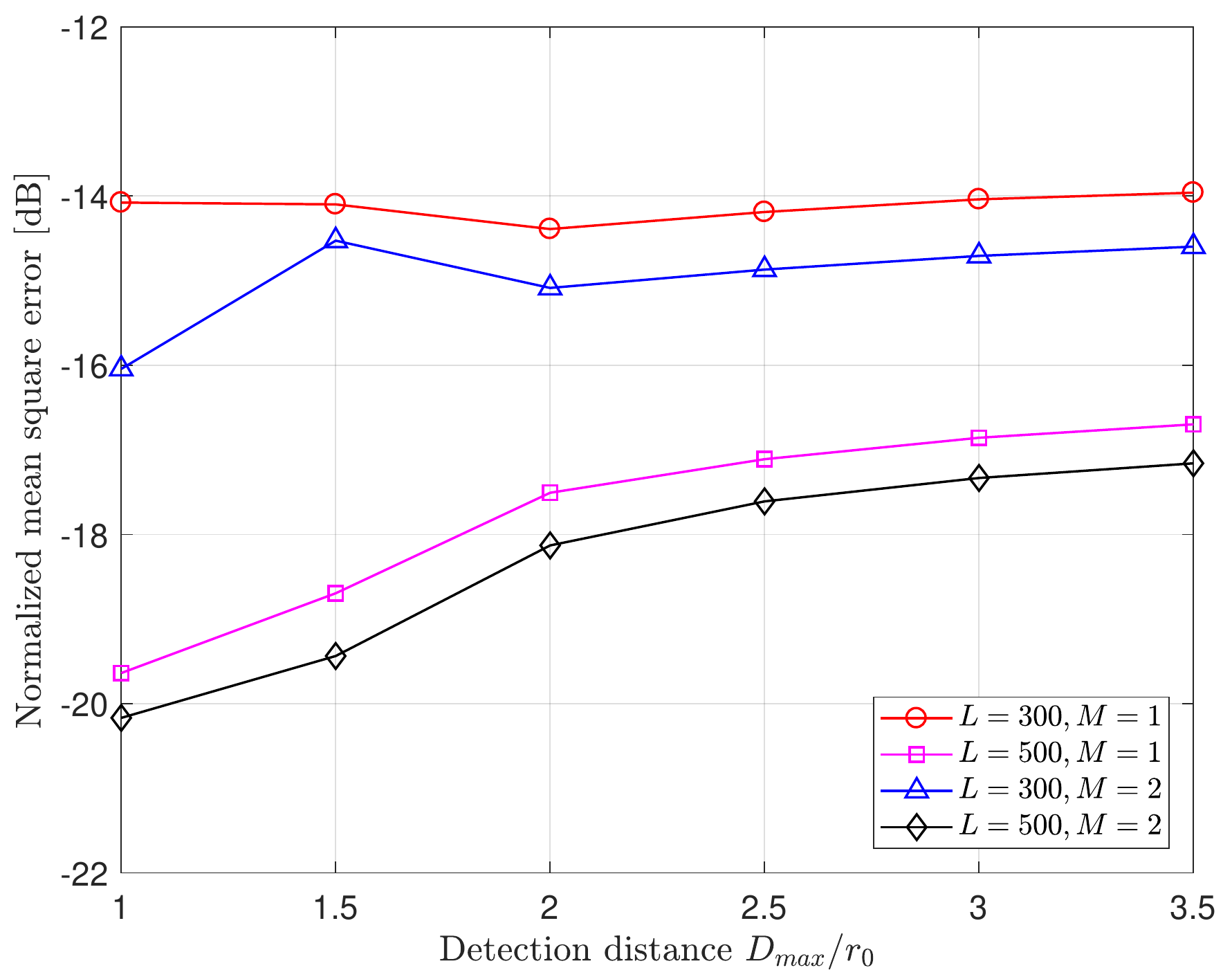}
    \vspace{-0.2cm}
    \caption{The CE performance of DCS-MMV-GAMP under different detection distances.}\label{Fig:NMSE_Dth}
  \end{minipage}
\end{figure}

%

\subsection{Performance Evaluation of the Proposed Detection Strategy}

First, we evaluate the impact of the maximal detection distance $D_{max}$ on the performance of AD and CE in Fig. \ref{Fig:EDP_Dth} and Fig. \ref{Fig:NMSE_Dth}, respectively. Here, we simply set $T_w = 4$, $M=1$, and $T_f = \{t_0+1,t_0+2\}$.
With $D_{max}$ being enlarged, each user is possible to be detected by more APs and more performance improvement can be provided by the AP cooperation. Thus, the user activity can be more accurately detected at the CU, which is validated in Fig. \ref{Fig:EDP_Dth}. It is observed that the proposed algorithm will have saturated performance if $D_{max}$ is large enough, meaning that we can set a moderate value of $D_{max}$ to save the computational cost. However, the CE performance is usually deteriorated by enlarging $D_{max}$ in Fig. \ref{Fig:NMSE_Dth}, which may be caused by the poor accuracy when estimating the small channel coefficients between the distant users and the APs. By accounting both the performance and complexity, we set $D_{max} = 2.5r_0$ in the rest of the simulations.

\begin{figure}[t]
  \centering
  \begin{minipage}[t]{.5\textwidth}
    \center
    \includegraphics[width=0.9\textwidth]{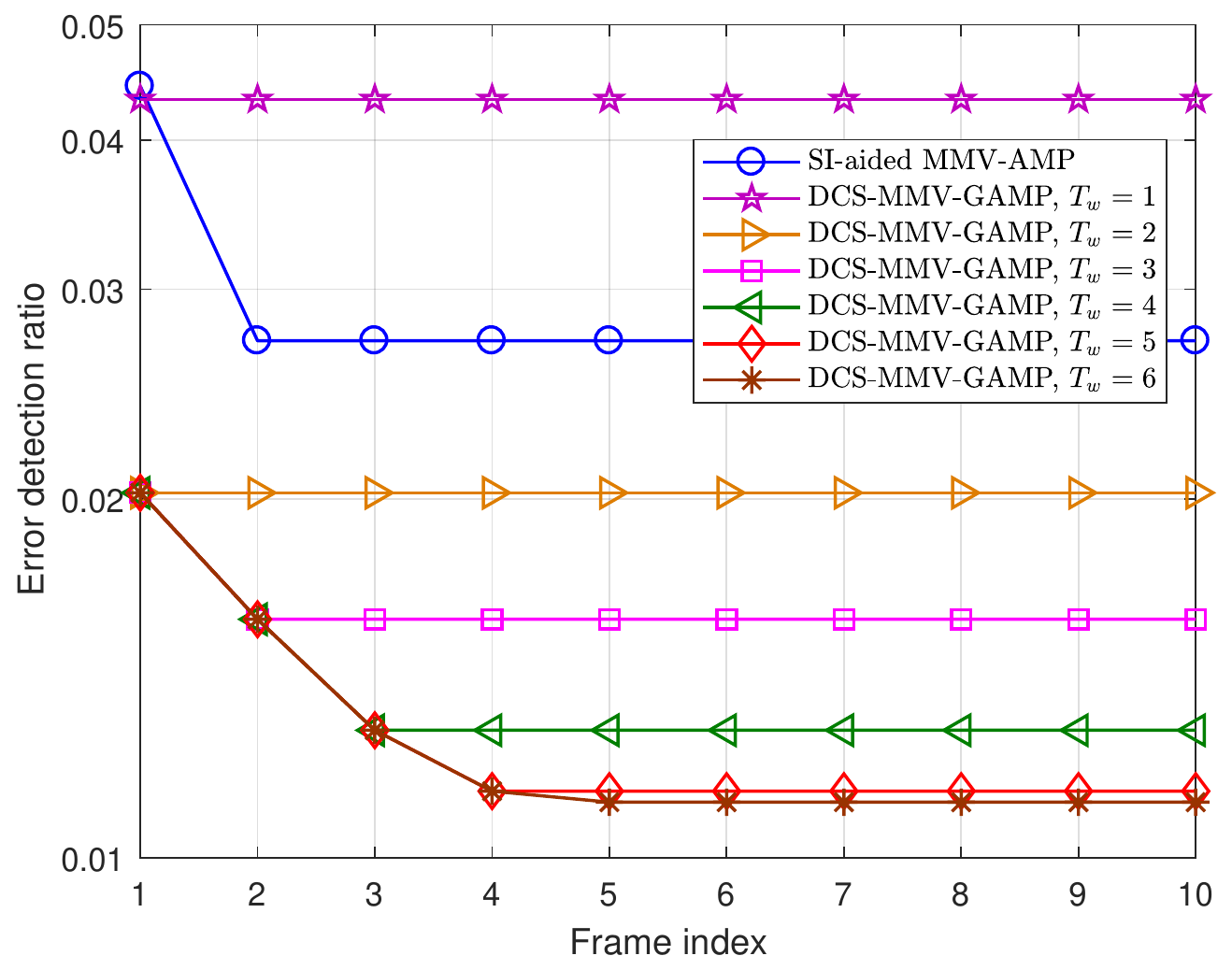}
    \vspace{-0.2cm}
    \caption{The AD performance under the detection window with different size when $M=1$.}\label{Fig:EDP_Tt}
    \vspace{0.3cm}
  \end{minipage}
  \begin{minipage}[t]{.5\textwidth}
    \center
    \includegraphics[width=0.9\textwidth]{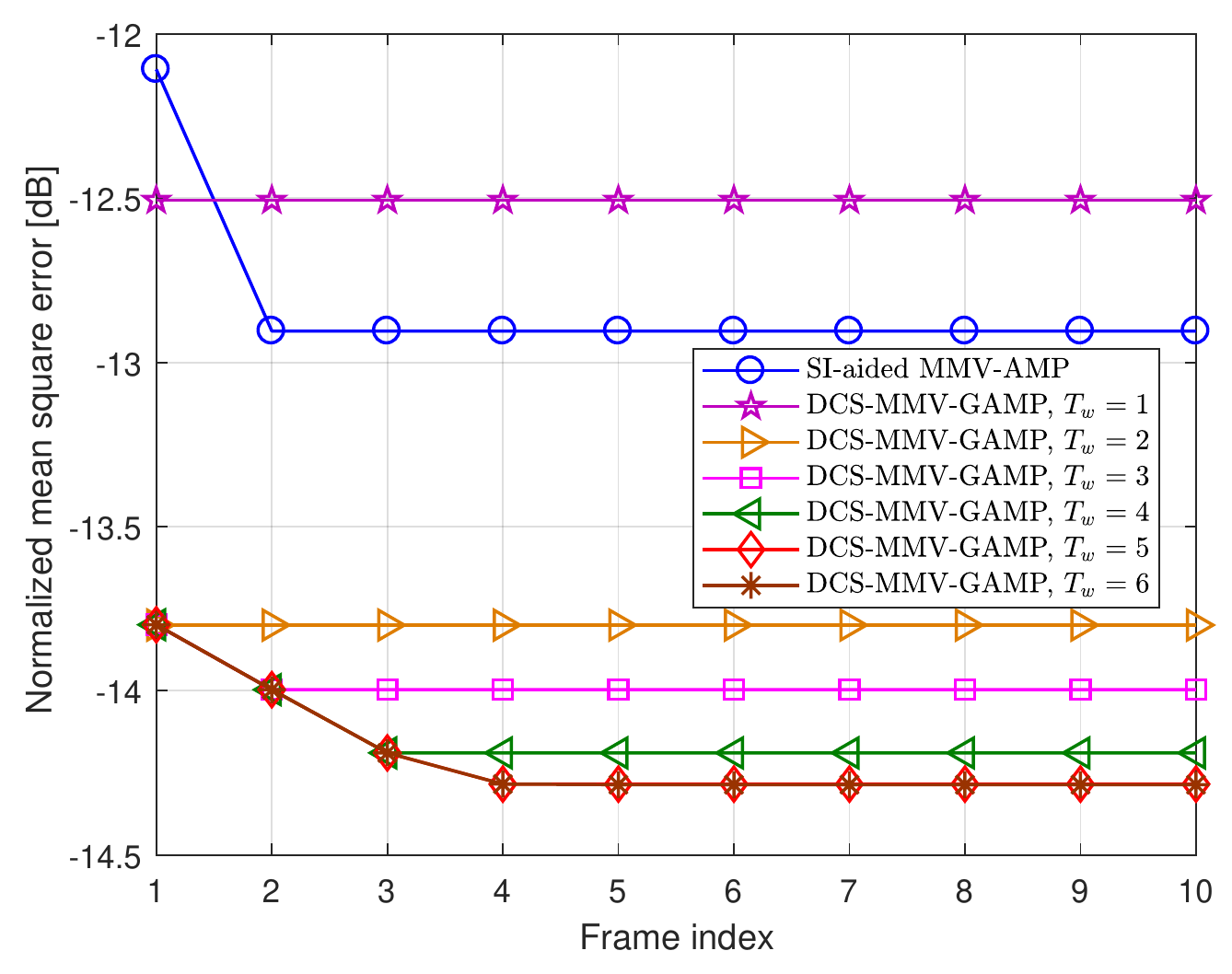}
    \vspace{-0.2cm}
    \caption{The CE performance under the detection window with different size when $M=1$.}\label{Fig:NMSE_Tt}
  \end{minipage}
\vspace{-0.7cm}
\end{figure}

%

Then we study the impact of the length of the detection window on the AD and CE performance with $L = 300, M=1$. The EDR and the NMSE in each frame of the $T=10$ frames are given in Fig. \ref{Fig:EDP_Tt} and Fig. \ref{Fig:NMSE_Tt}, respectively. Note that DCS-MMV-GAMP cannot employ the detection window with size $T_w$ before the detection for the $(T_w-1)$th frame.
For the special case with $T_w = 1$, DCS-MMP-GAMP reduces to CS-MMV-GAMP and thus has the same AD and CE performance in each frame.
In particular, the AD and CE performance of the first frame by DCS-MMV-GAMP is even better than that realized by the SI-aided MMV-AMP algorithm when $T_w \ge 2$, since the generalized sliding-window detection strategy enables us to exploit the temporal correlation between the activities in the adjacent frames. As $T_w$ increases, the DCS-MMV-GAMP algorithm will also have saturated performance, meaning that the proposed algorithm with a moderate $T_w$ can nearly achieve its optimal performance. To cope with the trilemma among performance, complexity, and latency, we set $T_w=4$ in the rest of the simulations.

\subsection{Performance Comparison under Different System Settings}

\begin{figure}[t]
  \centering
  \begin{minipage}[t]{.5\textwidth}
    \center
    \includegraphics[width=0.9\textwidth]{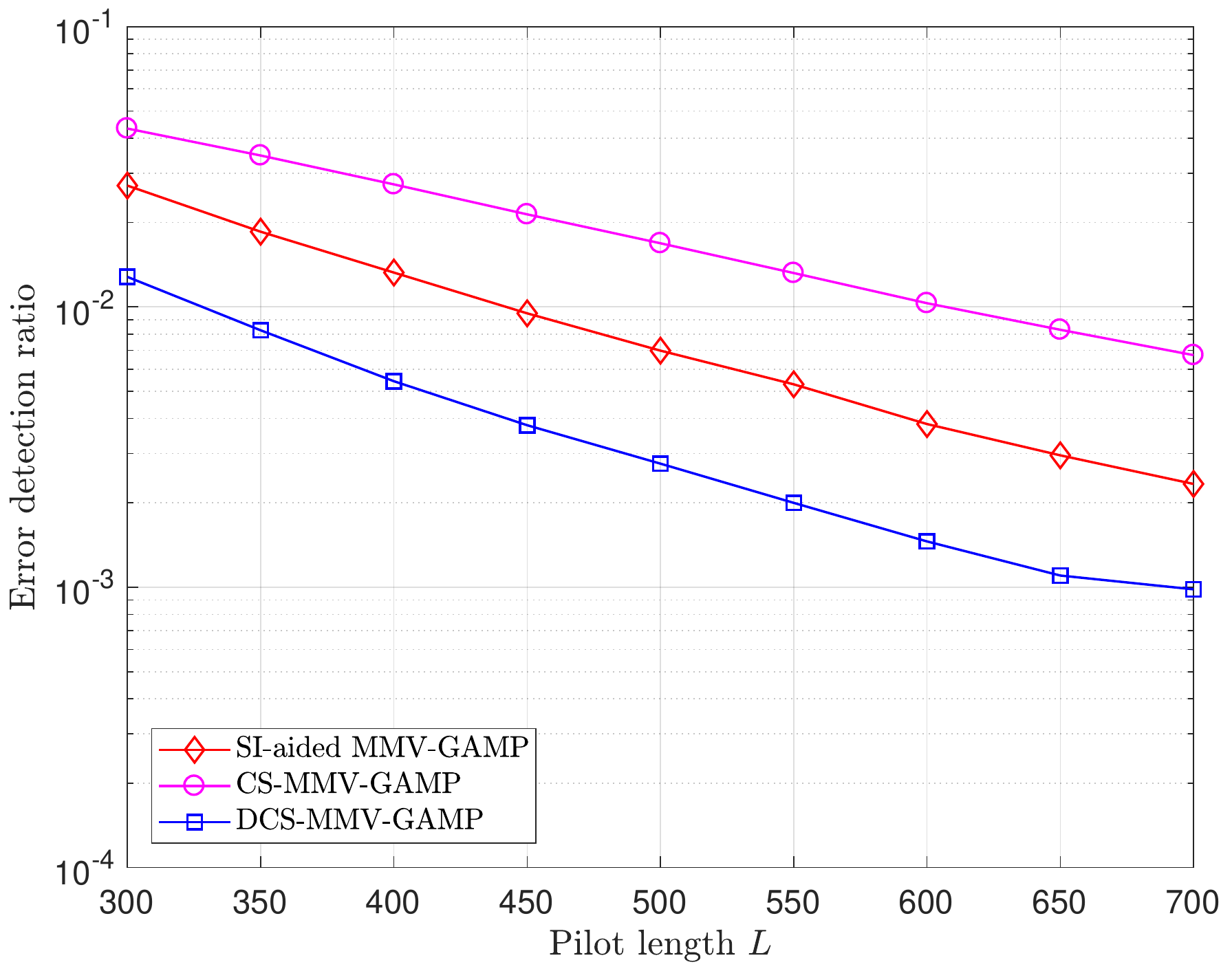}
    \vspace{-0.2cm}
    \caption{The impact of the pilot length on the AD performance with $M=1$.}\label{Fig:EDP_L}
    \vspace{0.3cm}
  \end{minipage}
  \begin{minipage}[t]{.5\textwidth}
    \center
    \includegraphics[width=0.9\textwidth]{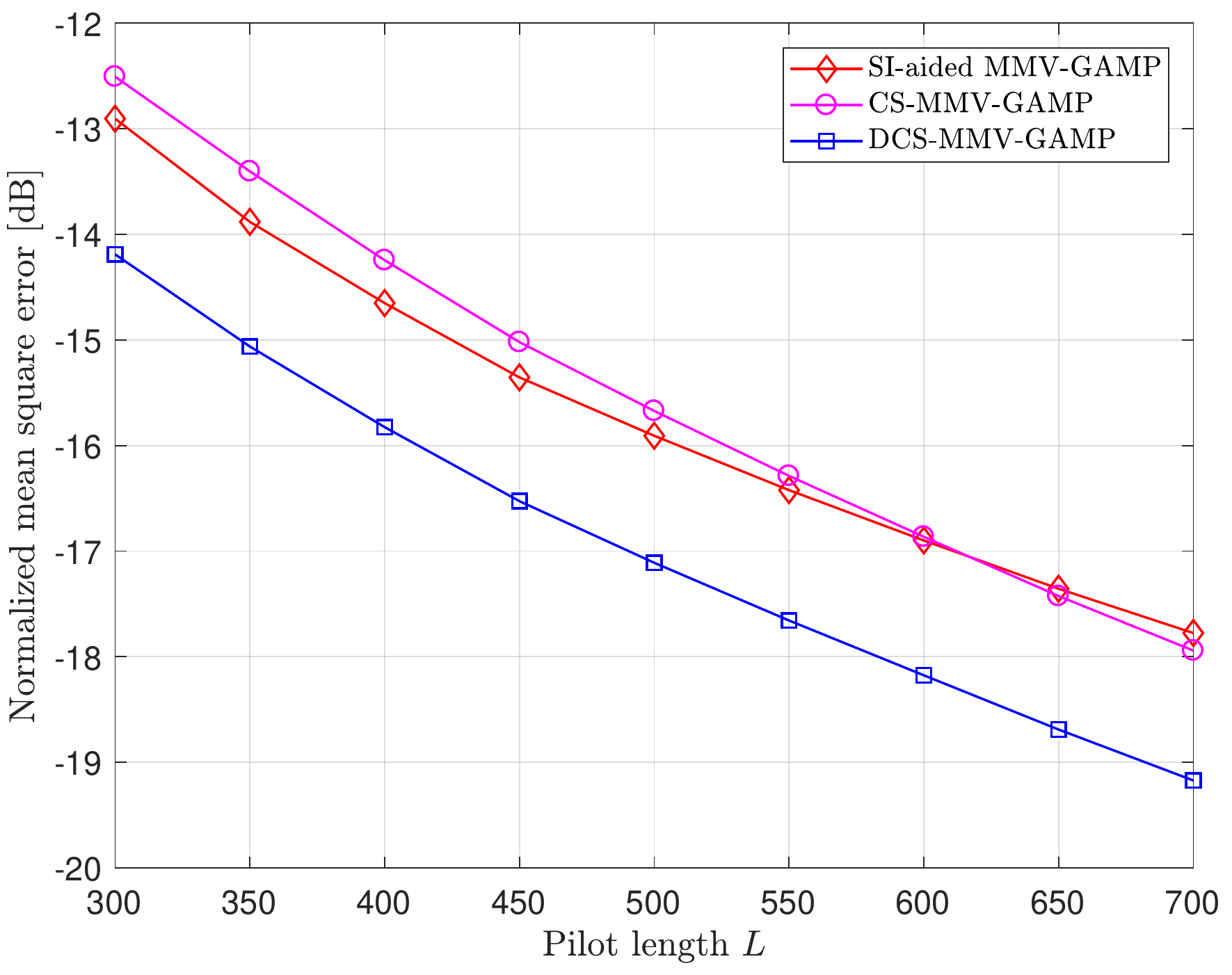}
    \vspace{-0.2cm}
    \caption{The impact of the pilot length on the CE performance with $M=1$.}\label{Fig:NMSE_L}
  \end{minipage}
\end{figure}

%

The performance comparison of DCS-MMV-GAMP and the benchmarks under different $L$ is given in Fig. \ref{Fig:EDP_L} and Fig. \ref{Fig:NMSE_L}. It is observed that both the AD and CE performance can be enhanced by increasing $L$. Though the NMSE of CS-MMV-GAMP becomes lower than that of SI-aided MMV-AMP when $L>650$, the AD performance of the CS-MMV-GAMP algorithm is always inferior to these two algorithms that exploit the temporal correlation. With $L$ increases, the proposed DCS-MMV-GAMP algorithm keeps performing much better than the benchmarks by introducing higher computational complexity to fully exploit the temporal correlation.

\begin{figure}[t]
  \centering
  \begin{minipage}[t]{.5\textwidth}
    \center
    \includegraphics[width=0.9\textwidth]{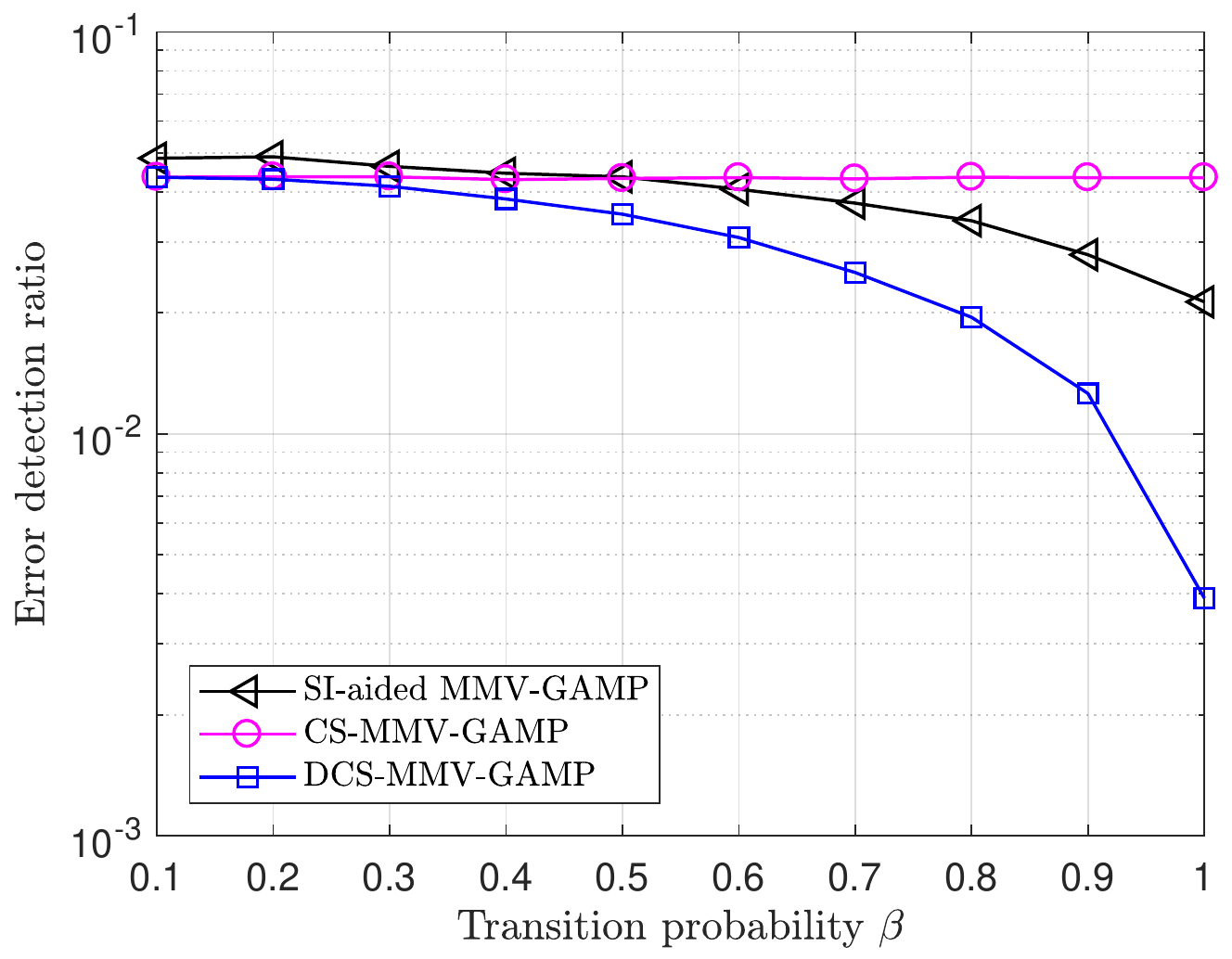}
    \vspace{-0.2cm}
    \caption{The impact of the transition probability $\beta$ on the AD performance with $p_a=0.1, M=1$.}\label{Fig:EDP_beta}
    \vspace{0.3cm}
  \end{minipage}
  \begin{minipage}[t]{.5\textwidth}
    \center
    \includegraphics[width=0.9\textwidth]{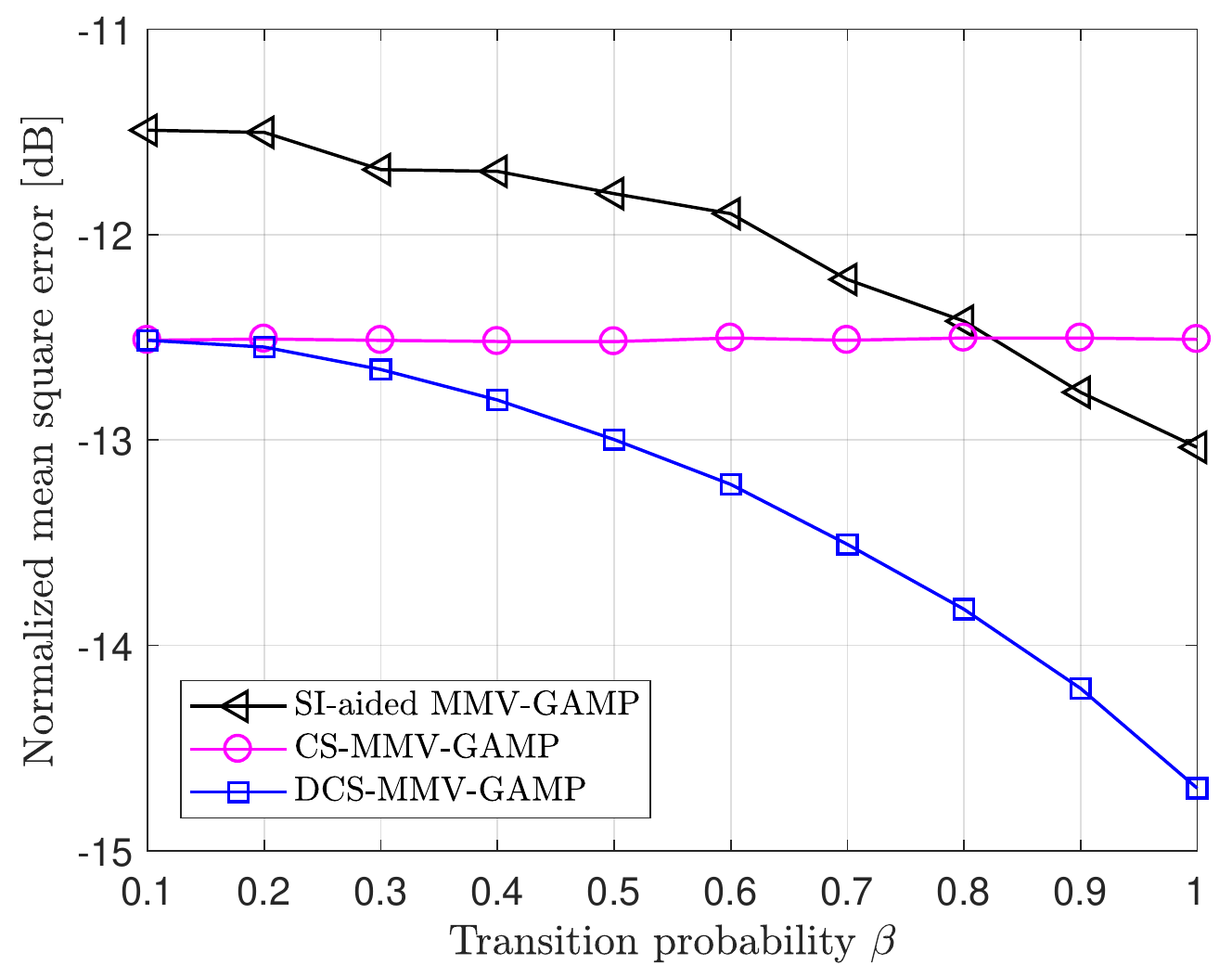}
    \vspace{-0.2cm}
    \caption{The impact of the transition probability $\beta$ on the CE performance with $p_a=0.1, M=1$.}\label{Fig:NMSE_beta}
  \end{minipage}
\end{figure}

%

\begin{figure}[t]
  \centering
  \begin{minipage}[t]{.5\textwidth}
    \center
    \includegraphics[width=0.9\textwidth]{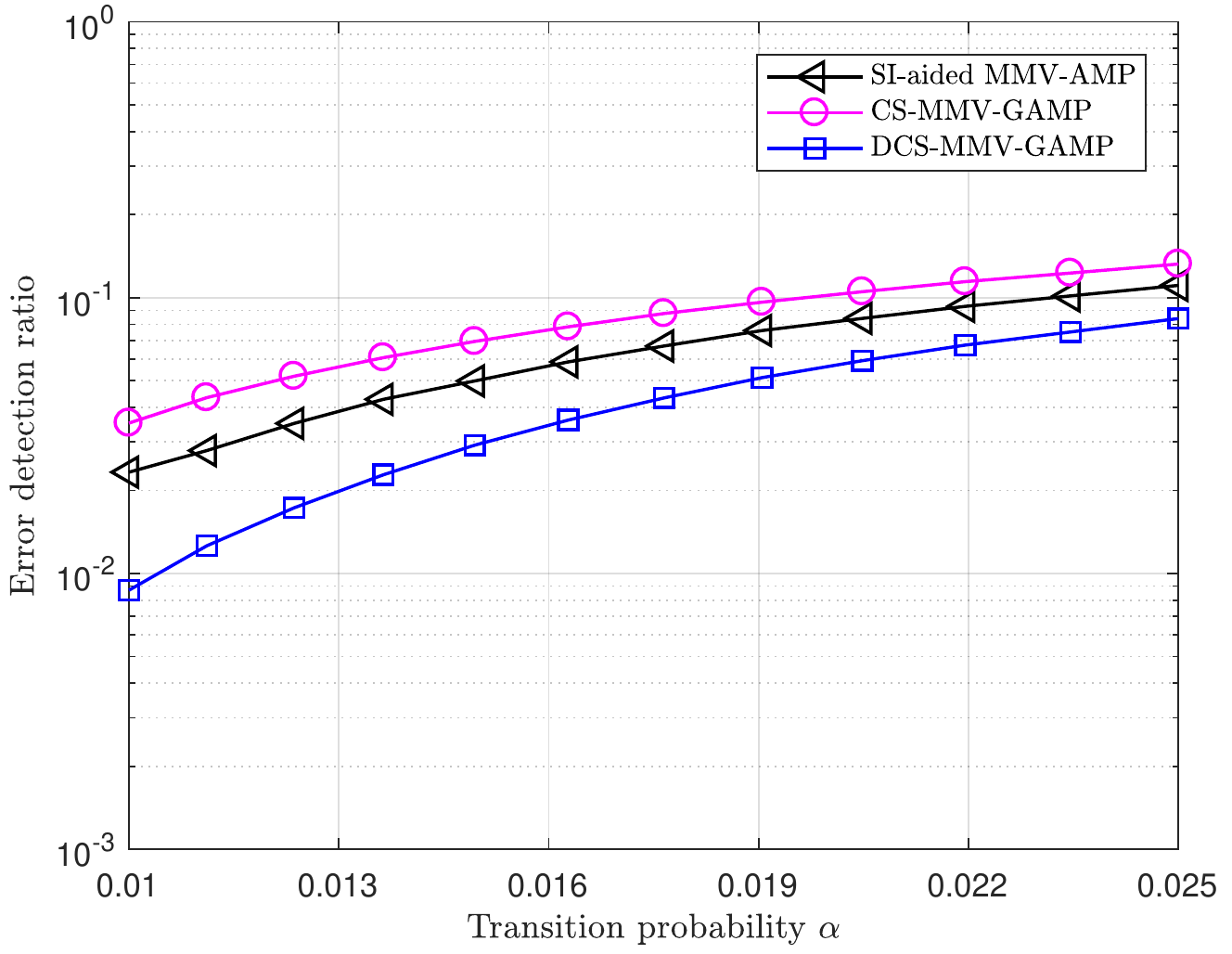}
    \vspace{-0.2cm}
    \caption{The impact of the transition probability $\alpha$ on the AD performance with $\beta=0.9, M=1$.}\label{Fig:EDP_alpha}
    \vspace{0.3cm}
  \end{minipage}
  \begin{minipage}[t]{.5\textwidth}
    \center
    \includegraphics[width=0.9\textwidth]{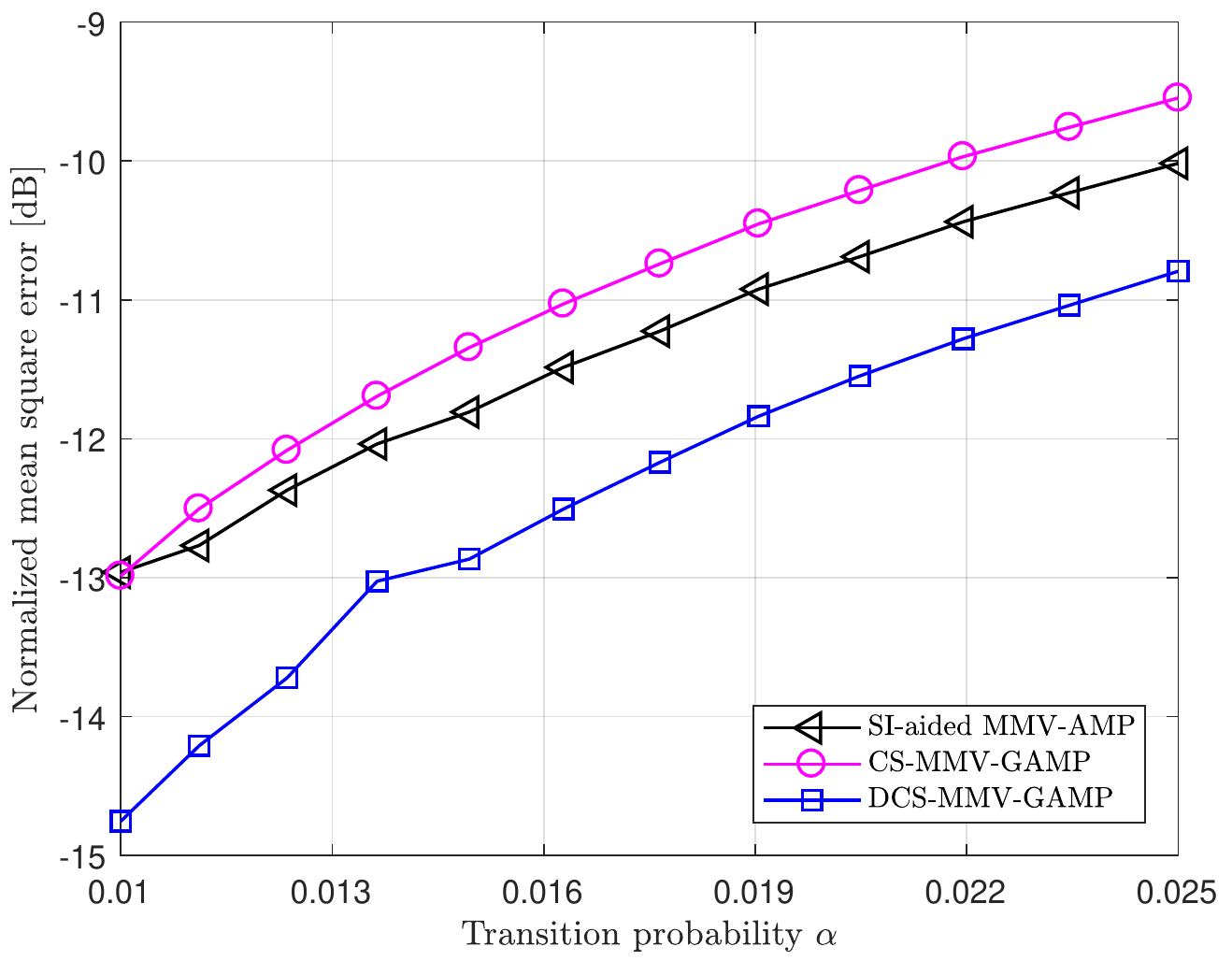}
    \vspace{-0.2cm}
    \caption{The impact of the transition probability $\alpha$ on the CE performance with $\beta=0.9, M=1$.}\label{Fig:NMSE_alpha}
  \end{minipage}
\end{figure}

%

The comparison of the proposed algorithm with the benchmarks under different values of transition probability $\beta$ is shown in Fig. \ref{Fig:EDP_beta} and Fig. \ref{Fig:NMSE_beta}, where $L = 300, p_a = 0.1, M=1$. It is seen that the proposed DCS-MMV-GAMP algorithm can always outperform the benchmarks when the user activity is temporally correlated, i.e., $\beta > p_a = 0.1$. When $\beta < 0.5$, the SI-aided MMV-AMP algorithm is even worse than the CS-MMV-GAMP algorithm, probably because the SI from the previous frame is quite inaccurate under a small pilot length and deteriorates the detection performance if the temporal correlation is not strong enough. As $\beta$ increases, the performance gaps between the DCS-MMV-GAMP algorithm and the benchmarks are also enlarged. We further evaluate the performance of the proposed algorithm under various values of transition probability $\alpha$ in Fig. \ref{Fig:EDP_alpha} and Fig. \ref{Fig:NMSE_alpha} with $L = 300, M=1, \beta = 0.9$. It is shown that the DCS-MMV-GAMP algorithm and the benchmarks all have degraded performance with larger $\alpha$.
As $\alpha$ increases, the active probability $p_a$ is also enlarged under fixed $\beta$ and thus more users are likely to be active in each frame. Then the AD and CE performance is deteriorated by the larger inter-user interference. It is noted that DCS-MMV-GAMP can always outperform the benchmarks, validating the superiority of the proposed algorithm.

\begin{figure}[t]
  \centering
  \includegraphics[width=.45\textwidth]{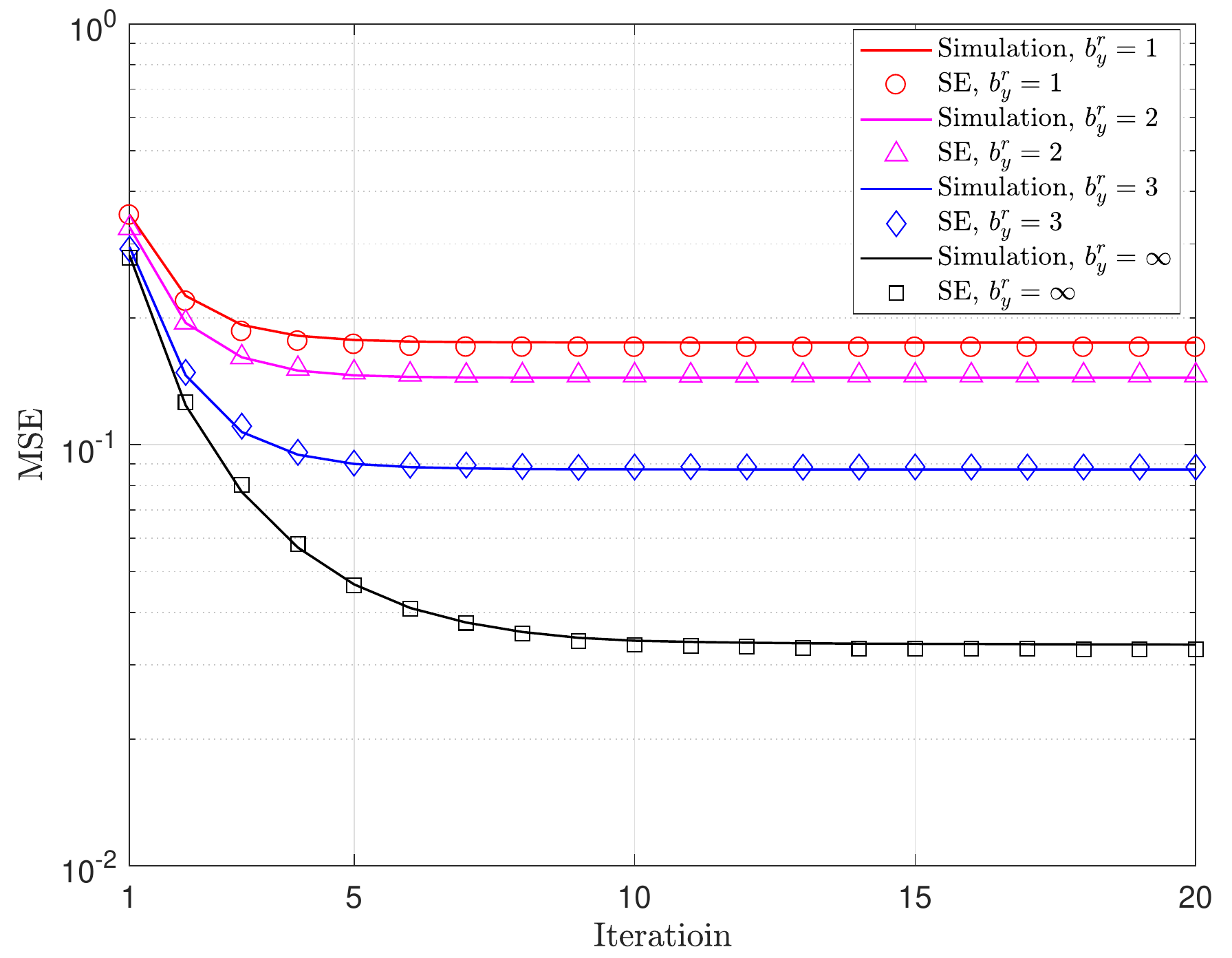}
  \caption{The convergence of DCS-MMV-GAMP.}\label{Fig:Conv}
\end{figure}
The convergence of the proposed DCS-MMV-GAMP algorithm is evaluated in Fig. \ref{Fig:Conv}. We set $L=300$ and $M=1$. It is observed that the performance of DCS-MMV-GAMP can be accurately predicted by the SE in both the infinite-fronthaul-capacity scenario and the finite-fronthaul-capacity scenario.
This result also indicates that we can utilize the SE to decide the system parameters before the specific system deployment.

%

\subsection{Performance Evaluation with Finite Fronthaul Capacity}

\begin{figure}[t]
  \centering
  \begin{minipage}[t]{.5\textwidth}
    \center
    \includegraphics[width=0.9\textwidth]{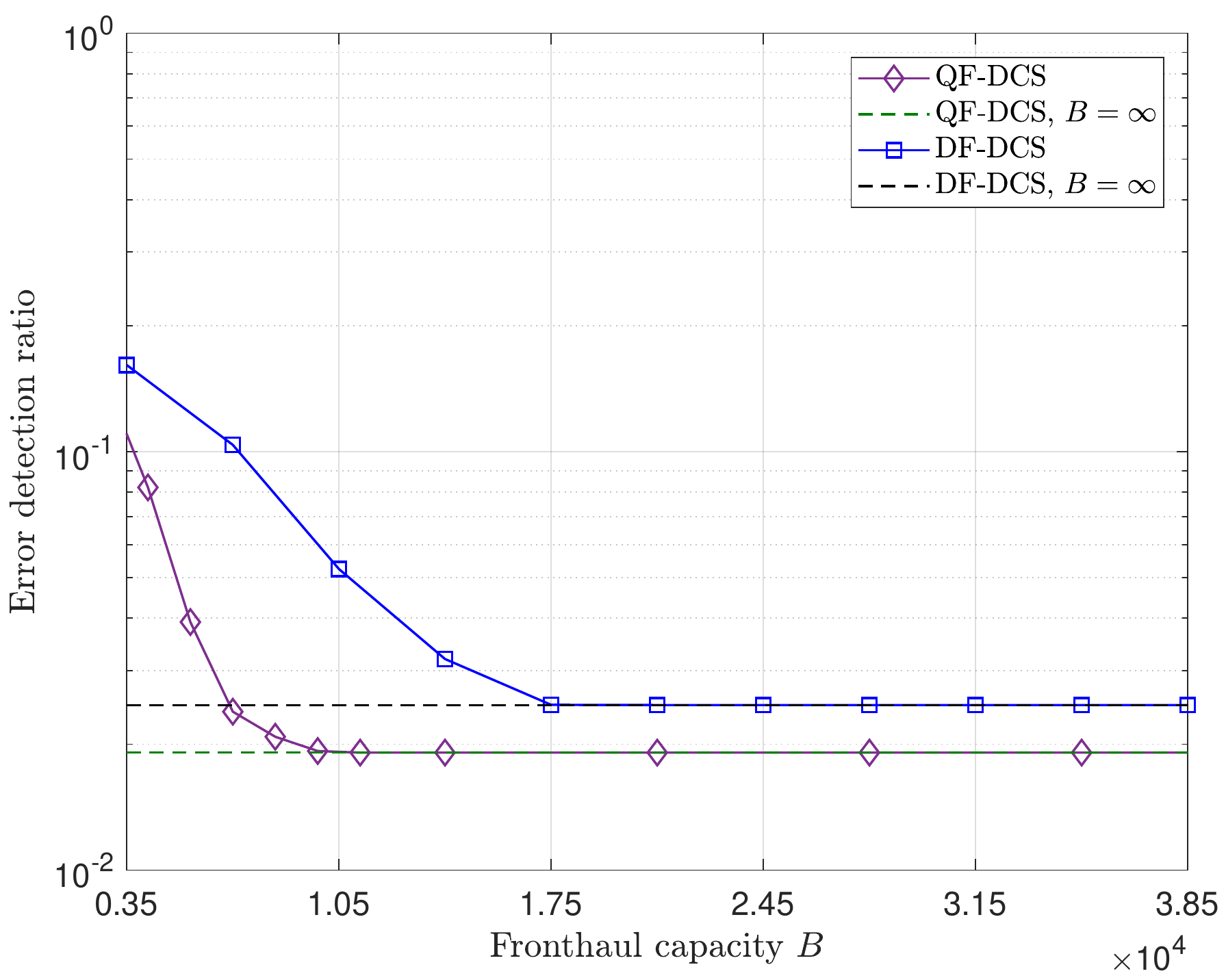}
    \vspace{-0.2cm}
    \caption{The comparison of QF and DF under different fronthaul capacities with $M=2$.}\label{Fig:EDP_B_M2}
    \vspace{0.3cm}
  \end{minipage}
  \begin{minipage}[t]{.5\textwidth}
    \center
    \includegraphics[width=0.9\textwidth]{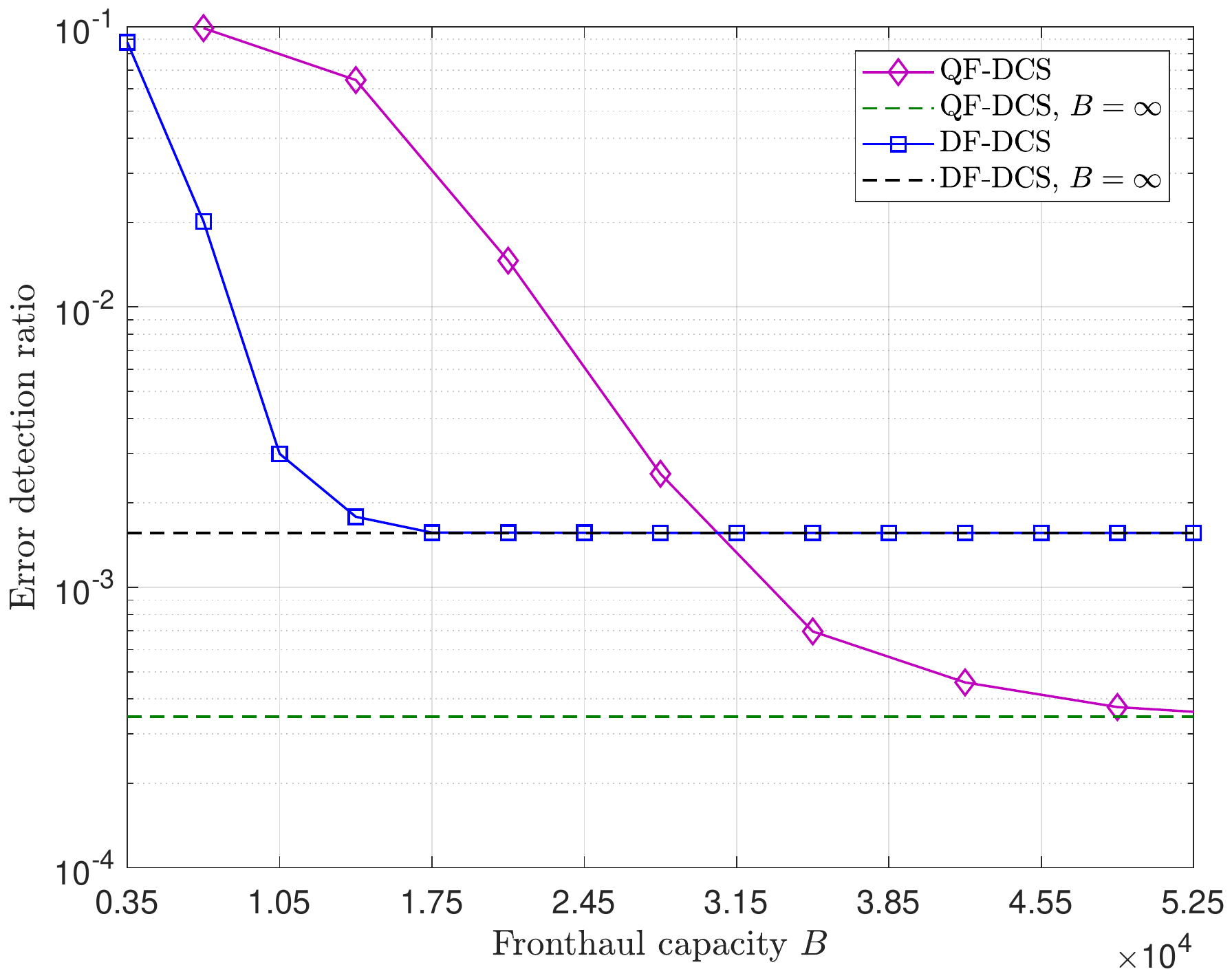}
    \vspace{-0.2cm}
    \caption{The comparison of QF and DF under different fronthaul capacities with $M=10$.}\label{Fig:EDP_B_M10}
  \end{minipage}
\end{figure}

The performance of QF and DF for the scenario with finite fronthaul capacity is evaluated under the system settings with different numbers $M$ of antennas in Fig. \ref{Fig:EDP_B_M2} and Fig. \ref{Fig:EDP_B_M10}. We compare these two schemes based on the proposed DCS-MMV-GAMP algorithm, which are denoted by QF-DCS and DF-DCS. Here, we set $L=350, N_c=500$ and $p_a=0.3$, then each AP detects the $7N_c = 3500$ users in its cell and the adjacent six cells for cooperation. First, in both the cases of $M=2$ and $M=10$, we observe that QF-DCS with $b^r_y=7$ can nearly achieve optimal performance without signal quantization, while $b_D=5$ is sufficient for DF-DCS to achieve the optimal performance. When $M=2$, QF-DCS always outperforms DF-DCS since the resolution of each sample in QF-DCS is much larger than that in DF-DCS.
When $M=10$, QF-DCS starts at $B=7000$ due to the fact that $B=2LMb^r_y \ge 7000$. Compared with QF-DCS, DF-DCS gives a dramatically larger reduction of the AD error when the fronthaul capacity is limited. Thus, DF-DCS is preferable for the system where APs are equipped with large-scale antenna arrays.


\section{Conclusion}\label{sec:conclusion}

This paper studies the cooperative AD and CE framework for massive access with temporally correlated activity in multi-cell networks.
We consider the user-centric AP cooperation strategy for complexity reduction and the generalized sliding window detection strategy for the exploit of the temporal correlation. In particular, a scalable DCS-MMV-GAMP algorithm with approximately Bayes-optimal performance is proposed, which can fully exploit the spatial-temporal correlation in activities. Furthermore, we develop two schemes of QF and DF to realize cooperative AD in practical systems with limited fronthaul capacity, where the impact of the antenna number on their performance is extensively investigated.
Numerical results demonstrate that the proposed DCS-MMV-GAMP algorithm can have greatly superior performance to the existing methods.
Moreover, we show that QF is usually favorable in the scenario where each AP has only a few antennas, while DF can significantly outperform QF with limited fronthaul capacity when $M$ is large.

\bibliographystyle{IEEEtran}
\bibliography{CFTCMC}

\end{document}